\documentclass[fleqn,usenatbib]{mnras}

\usepackage{newtxtext,newtxmath}

\usepackage[T1]{fontenc}

\DeclareRobustCommand{\VAN}[3]{#2}
\let\VANthebibliography\thebibliography
\def\thebibliography{\DeclareRobustCommand{\VAN}[3]{##3}\VANthebibliography}

\usepackage{graphicx}	
\usepackage{amsmath}	
\usepackage[dvipsnames]{xcolor}

\newcommand{\hMsun}{h^{-1}\mathrm{M_\odot}}
\newcommand{\hMpc}{h^{-1}\mathrm{Mpc}}
\newcommand{\hGpc}{h^{-1}\mathrm{Gpc}}
\newcommand{\hkpc}{h^{-1}\mathrm{kpc}}

\newcommand{\kms}{\mathrm{km\,s^{-1}}}

\newcommand{\sqdeg}{\mathrm{deg}^2}
\newcommand{\persqdeg}{\mathrm{deg}^{-2}}

\newcommand{\magr}{{}^{0.1}M_r}
\newcommand{\gr}{{}^{0.1}(g-r)}
\newcommand{\kcorrr}{{}^{0.1}k_r}
\newcommand{\kcorrg}{{}^{0.1}k_g}

\title[MXXL lightcone catalogue]{A lightcone catalogue from the Millennium-XXL simulation: improved spatial interpolation and colour distributions for the DESI BGS}

\author[A. Smith et al.]{
Alex Smith$^{1,2,3}$\thanks{E-mail: alex.smith@ed.ac.uk},
Shaun Cole$^{1}$,
Cameron Grove$^{1}$,
Peder Norberg$^{1,4}$,
and Pauline Zarrouk$^{5}$
\vspace*{4pt} \\ 
\scriptsize $^{1}$ Institute for Computational Cosmology, Department of Physics, Durham University, South Road, Durham DH1 3LE, UK\vspace*{-2pt}\\
\scriptsize $^{2}$ IRFU, CEA, Universit\'e Paris-Saclay, F-91191 Gif-sur-Yvette, France\vspace*{-2pt} \\
\scriptsize $^{3}$ Institute for Astronomy, University of Edinburgh, Royal Observatory, Blackford Hill, Edinburgh EH9 3HJ, UK\vspace*{-2pt}\\
\scriptsize $^{4}$ Centre for Extragalactic Astronomy, Department of Physics, Durham University, South Road, Durham DH1 3LE, UK \vspace*{-2pt} \\
\scriptsize $^{5}$ Sorbonne Universit\'e, Universit\'e Paris Diderot, CNRS/IN2P3, Laboratoire de Physique Nucl\'eaire et de Hautes Energies, LPNHE, 4 place Jussieu, F-75252 Paris, France \vspace*{-2pt} \\
}

\date{Accepted XXX. Received YYY; in original form ZZZ}

\pubyear{2022}

\begin{document}
\label{firstpage}
\pagerange{\pageref{firstpage}--\pageref{lastpage}}
\maketitle

\begin{abstract}
The use of realistic mock galaxy catalogues is essential in the preparation of large galaxy
surveys, in order to test and validate theoretical models and to assess systematics. We present an updated
version of the mock catalogue constructed from the Millennium-XXL simulation, which uses a halo occupation distribution (HOD) method to assign galaxies $r$-band magnitudes and $g-r$ colours. We have made several
modifications to the mock to improve the agreement with measurements from the SDSS and GAMA surveys. We find
that cubic interpolation, which was used to build the original halo lightcone,
produces extreme velocities between snapshots. Using linear interpolation improves the correlation function quadrupole measurements on small scales.
We also update the $g-r$ colour distributions
so that the observed colours better agree with measurements from GAMA data, particularly
for faint galaxies. As an example of the science that can be done with the mock, we investigate how the 
luminosity function depends on environment and colour, and find good agreement with measurements
from the GAMA survey.
This full-sky mock catalogue is designed for the ongoing Dark Energy Spectroscopic Instrument (DESI)
Bright Galaxy Survey (BGS), and is complete to a magnitude limit $r=20.2$.
\end{abstract}




\section{Introduction}

Over the past few decades, the large-scale structure of the Universe has been probed through
the use of large photometric and spectroscopic galaxy surveys. The two-point clustering statistics
of galaxies in these surveys allows us to measure the expansion history and growth rate of structure, by
taking baryon acoustic oscillation \citep[BAO; e.g.][]{Cole2005,Eisenstein2005} and 
redshift space distortion \citep[RSD;][]{Kaiser1987} measurements.
Targeting different types of galaxies enables us to make these measurements over a range of different
redshifts, placing constraints on models of dark energy, and testing general relativity \citep[e.g.][]{Guzzo2008}.
The increasing size of these surveys enables these constraints to be improved over time.

The Dark Energy Spectroscopic Instrument \citep[DESI;][]{DESI2016science,DESI2016instrument,DESI2022} has begun to conduct a survey of over 30 million
galaxies, targeting Luminous Red Galaxies (LRGs), Emission Line Galaxies (ELGs) and quasars (QSOs), which cover a wide
range of redshifts. During bright time, when the sky is bright due to the moon phase or twilight,
a survey of bright, nearby galaxies is being conducted, called the 
Bright Galaxy Survey (BGS). The BGS survey will target 10 million
galaxies in the primary BGS BRIGHT sample. This is a flux limited sample with $r$-band magnitude $r<19.5$, and density $\sim 800~\persqdeg$. In addition to this, there is a
secondary BGS FAINT sample, which extends the BGS to fainter magnitudes $19.5<r<20.175$,
with additional cuts based on colour and fibre magnitude to ensure a high
redshift success rate \citep{Hahn2022}.\footnote{The definition of the BGS FAINT sample has been updated since \citet{Ruiz2021} and \citet{Zarrouk2021}, where previously it was $19.5<r<20.0$.}
This lower priority sample will have a density of $\sim 600~\persqdeg$.
Target BGS galaxies are identified from the photometry of the DESI Legacy Imaging Surveys 
\citep{Dey2019,Ruiz2021}. This is made up of the Dark Energy Camera Legacy Survey 
(DECaLS), the Beijing-Arizona Sky Survey (BASS), and the Mayall z-band Legacy Survey
(MzLS), which together cover the $14\,000~\sqdeg$ footprint of the DESI survey.

In order for DESI to achieve the high precision measurements required to tighten
constraints on models of dark energy and gravity, it is essential that realistic mock
catalogues are used. Since the true cosmology of the simulation is known, 
mock catalogues can be used to test and validate the theoretical models used in the 
clustering analysis, and to quantify and correct for systematic effects. For example,
this was done in \citet{Smith2020} for the QSO sample of the  
extended Baryon Oscillation Spectroscopic Survey \citep[eBOSS;][]{Dawson2016},
with similar mock challenges for the LRG \citep{Rossi2021} and ELG samples \citep{Alam2021}. 

Since current galaxy surveys probe such large volumes, the N-body simulations used
to generate mock catalogues typically only contain dark matter, with galaxies subsequently
being placed within the dark matter haloes. There are several methods which can be used to populate dark matter haloes with galaxies. This includes the halo occupation
distribution \citep[HOD; e.g.][]{Peacock2000,Berlind2002,Zheng2005,Zehavi2011,Smith2017,Smith2020,Alam2021,Rossi2021}, which describes the probability that a halo of mass $M_\mathrm{h}$ contains
central and satellite galaxies. The central galaxies are then placed at the
centre of the halo, with satellites positioned using the simulation particles, or
following an analytic profile. Subhalo abundance matching \citep[SHAM; e.g.][]{Vale2004,Conroy2006,Rodriguez2016,Safonova2021} ranks subhaloes
based on a property such as halo mass or circular velocity. Galaxies are placed
within each subhalo, and assigned luminosities or stellar masses
based on the subhalo ranking (with scatter). Semi-analytic models \citep[SAM; e.g.][]{Cole2000,Benson2010} solve
a set of differential equations which model the physics of galaxy formation and evolution.

To create a realistic mock catalogue that mimics a real galaxy 
survey, galaxies need to be positioned within a simulated lightcone.
On-the-fly lightcone outputs are available for some simulations \citep[e.g.][]{Potter2017, Maksimova2021}, but are uncommon; most simulation outputs are in the periodic cubic box, 
at discrete snapshot times.
Approximate lightcones are commonly constructed from the snapshots by simply splicing them together in 
spherical shells \citep[e.g.][]{Fosalba2008, Giocoli2016, Avila2018, Comparat2019, DongPaez2022, Wang2022}. 
However, this method leads to discontinuities in the lightcone,
and it is possible for the same halo to be replicated 
\citep[for a detailed discussion on these issues, see][]{Smith2022}. 
Alternatively, if halo merger trees are available, interpolation can be used.

A mock catalogue was previously made in \citet{Smith2017} from the Millennium-XXL
simulation, which was designed for the DESI BGS.
A halo lightcone was first constructed by interpolating haloes between simulation 
snapshots, finding the time at which they cross the observer's lightcone. The 
halo lightcone was made to $z=2.2$, making it useful for a range of future
galaxy surveys. The halo lightcone was then populated with galaxies to create
a BGS mock, using a HOD scheme. A set of nested HODs for different magnitude thresholds
was used to assign galaxies in the mock an $r$-band magnitude, and each galaxy was 
also assigned a $g-r$ colour. The luminosity function and colour
distributions were tuned to reproduce measurements from the 
Sloan Digital Sky Survey \citep[SDSS;][]{Abazajian2009} and Galaxy and Mass Assembly
(GAMA) survey \citep{Driver2009,Driver2011,Liske2015}.

The MXXL halo lightcone used cubic interpolation to interpolate positions and velocities
between simulation snapshots. Cubic interpolation was also previously used to make a SAM galaxy lightcone 
in \citet{Merson2013}, from the Millennium simulation \citep{Springel2005}.
Other examples where linear interpolation was used include \citet{Izquierdo2019} and \citet{Korytov2019}. In \citet{Hadzhiyska2022}, halo lightcones are provided from the
AbacusSummit simulation \citep{Maksimova2021} both with linear interpolation, and by
matching snapshot haloes to the particle lightcone outputs.

The MXXL mock has been widely used in the preparation of the BGS, and the halo
catalogue has been used to create mock catalogues for other surveys.
The galaxy mock was used to quantify the effect of DESI fibre incompleteness on galaxy clustering in
\citet{Smith2019}, and assess correction methods. The clustering statistics of
the mock were compared with BGS targets in the Legacy Imaging Survey DR9 in \citet{Zarrouk2021}.
ELG mocks for Euclid \citep{Laureijs2011} and the Roman Space Telescope \citep{Spergel2015} surveys were made from the halo lightcone, using HODs
from the Galacticus SAM in \citet{Merson2019}, which were used to make linear bias forecasts.
QSO mocks were made in \citet{Kovacs2021}, to study the integrated Sachs-Wolfe (ISW) signal \citep{SachsWolfe1967}.

In this work, we make improvements to the original MXXL mock, by modifying how
the haloes in the lightcone are interpolated, and by improving the $g-r$
colour distributions. This mock will be useful for the mock challenges in DESI,
and will complement other mocks being created from other simulations, in 
a range of cosmologies, e.g. SHAM mocks from the Uchuu simulation \citep[][]{DongPaez2022,Ishiyama2021}
and HOD mocks from the AbacusSummit simulations \citep[][Grove et al., in prep]{Maksimova2021}. 

This paper is organised as follows. Section~\ref{sec:mxxl} describes the MXXL simulation
and gives an overview of the methods used to create the original MXXL mock. In Section~\ref{sec:interpolation},
we investigate different methods for interpolating haloes. 
Section~\ref{sec:colours} describes the improvement to the $\gr$ colour distributions in the mock, which are fit directly to GAMA measurements. In
Section~\ref{sec:lf_environment}, we investigate how the luminosity function in the
mock depends on environment and colour. Finally, we summarise our conclusions in
Section~\ref{sec:conclusions}.

\section{MXXL mock}
\label{sec:mxxl}

In this section, we give an overview of the MXXL halo lightcone and galaxy catalogue
previously created in \citet{Smith2017}.

\subsection{The MXXL simulation}
The Millennium-XXL (MXXL) simulation \citep{Angulo2012} is a dark-matter-only N-body simulation with box 
size $3~\hGpc$ and particle
mass $6.17\times10^9~\hMpc$, in a WMAP1 cosmology with $\Omega_\mathrm{m}=0.25$,
$\Omega_\Lambda=0.75$, $\sigma_8=0.9$, $h=0.73$ and $n_s=1$ \citep{Spergel2003}. 

Friends-of-friends \citep[FOF;][]{Davis1985} halo catalogues were output at multiple simulation snapshots.
Bound subhaloes were found using the \textsc{subfind} algorithm \citep{Springel2001}, and halo merger trees were
built by identifying the descendant of a halo at the next snapshot. This was 
done using the 15 most bound particles in each halo.

\subsection{MXXL halo lightcone}

The MXXL halo lightcone was constructed by interpolating halo properties between snapshots,
finding the redshift at which the interpolated haloes cross the observer's lightcone.
This was made possible by using the pre-computed halo merger trees, which allow the unique 
descendant of each halo to be easily identified. 
An observer was first placed at a random location in the cubic box, and halo
positions and velocities were interpolated using cubic interpolation, with the
initial and final positions and velocities as boundary conditions.
Halo masses (defined as $M_\mathrm{200m}$, the mass enclosed within a sphere with density 200 times
the mean density of the Universe), were interpolated linearly.

If two or more haloes have the same descendant, a merger takes place in the time between the two snapshots,
but it is not known when this occurs. 
When this happens, the total mass of all merging progenitor haloes is interpolated linearly in time, $t$,
with each progenitor assigned a constant fraction of the total mass. A random time
is chosen between the two snapshots for each merger to take place. If the progenitor halo
crosses the lightcone after this time, all of its mass is transferred to the most massive
progenitor.

The full-sky MXXL halo catalogue was constructed to a maximum redshift of $z=2.2$. 
To achieve this, the periodic box must be replicated for
haloes with $z \gtrsim 0.5$.

\subsection{MXXL galaxy catalogue}
\label{sec:mock_creation_method}

In \citet{Smith2017}, the halo catalogue is populated with galaxies to
an apparent magnitude limit of $r=20$. Here we summarise the method of creating the mock.
Each galaxy in the mock is assigned a
rest-frame $r$-band absolute magnitude, $k$-corrected to a reference
redshift $z_\mathrm{ref}=0.1$, which we denote as $\magr$.
For clarity, we use $h=1$ to drop the $5\log_{10} h$ term. 
Rest-frame colours, $\gr$, are also $k$-corrected to the same $z_\mathrm{ref}=0.1$. 
Observer-frame colours are denoted as $g-r$, without the superscript $0.1$.

\subsubsection{Halo occupation distribution}

The MXXL halo lightcone was populated with galaxies using a halo occupation distribution 
(HOD) scheme. This was done using a set of HODs for different $r$-band absolute magnitude 
thresholds, $\magr$, 
which were measured from the SDSS survey \citep{Zehavi2011}.

The HOD describes the average number of galaxies, brighter than luminosity $L$,
that reside in a halo as a function of halo mass, $M_\mathrm{h}$. This can be split
into a central galaxy at the centre of the halo, which is surrounded by satellites,
\begin{equation}
\langle N(>L|M_\mathrm{h}) \rangle = \langle N_\mathrm{cen}(>L|M_\mathrm{h}) \rangle + \langle N_\mathrm{sat}(>L|M_\mathrm{h}) \rangle.
\end{equation}
The central HOD is modelled as a smoothed step function,
\begin{equation}
\langle N_\mathrm{cen}(>L|M_\mathrm{h}) \rangle = \frac{1}{2}\left[1 + \mathrm{erf}\left(\frac{\log M_\mathrm{h} - \log M_\mathrm{min}(L)}{\sigma_{\log M}(L)} \right)\right],
\end{equation}
and the satellite HOD is a power law, which is multiplied by the central HOD to prevent
there being haloes with satellites brighter than the central galaxy,
\begin{equation}
\langle N_\mathrm{sat}(>L|M_\mathrm{h}) \rangle = \langle N_\mathrm{cen}(>L|M_\mathrm{h}) \rangle \left( \frac{M_\mathrm{h} - M_0(L)}{M_1(L)} \right)^{\alpha(L)}.
\end{equation}
In total, there are 5 HOD parameters, where $M_\mathrm{min}$ and $\sigma_{\log M}$ set the
position and width of the step function for centrals, while for satellites $M_1$ is the 
average halo mass with 1 satellite, $M_0$ is a low mass cutoff, and $\alpha$ is the
power law slope. For the central HOD, the error function, $\mathrm{erf}(x)$, is modified
to use a pseudo-Gaussian spline kernel function \citep[see equations 8-10 of][]{Smith2017}.
This is done to prevent unphysical crossing of the HODs for different magnitude thresholds.

Smooth functions were fit to each HOD parameter as a function of magnitude. 
These smooth curves allow the galaxies in the mock to be assigned 
absolute magnitudes. 
A random number from the pseudo-Gaussian function is generated for each central galaxy, 
which sets the location of the galaxy on the smooth step function. A root-finding procedure is then used to convert this to an absolute magnitude, finding the $L$ where 
$x\sqrt{2}\sigma_{\log M}(L) = \log M_\mathrm{h} - \log M_\mathrm{min}(L)$. A factor
of $\sqrt{2}$ is needed, due to how $\sigma_{\log M}$ is defined.
The total number of satellites in each halo above a faint magnitude limit, $L_\mathrm{min}$ is then calculated
by drawing a random number from a Poisson distribution, with mean $\langle N_\mathrm{sat}( > L_\mathrm{min} | M_\mathrm{h}) \rangle$. Each satellite is then assigned a magnitude by generating a uniform
random number $0 < u < 1$, and solving $\langle N_\mathrm{sat}( > L | M_\mathrm{h}) \rangle$ / $\langle N_\mathrm{sat}( > L_\mathrm{min} | M_\mathrm{h}) \rangle = u$. 
The satellites are positioned around the central galaxy following a NFW profile \citep{NFW1997}.
Since the HODs come from fits to the SDSS data, this reproduces the SDSS luminosity function at $z=0.1$.


\subsubsection{Evolution of the HODs}

To evolve the HODs with redshift, a target $r$-band luminosity function is first defined that
we aim to reproduce in the mock. The luminosity function used comes from SDSS
measurements at low redshift \citep{Blanton2003}, and transitions to a Schechter function fit 
to the GAMA luminosity function at higher redshifts \citep{Loveday2012}. 
This transition occurs at $z=0.15$.
The evolution parameters used are $P=1.8$ and $Q=0.7$, which describe the
evolution of number density and luminosity, respectively. The Schechter
parameters $M^*$ and $\phi^*$ are evolved using the $P$ and $Q$ parameters as
\begin{equation}
\begin{split}
M^*(z) &= M^*(z_0) - Q(z-z_0)\\
\phi^*(z) &= \phi^*(0)10^{0.4Pz},
\end{split}
\end{equation}
where $z_0=0.1$, $M^*$ is the characteristic magnitude at the exponential cut-off in the power law, and the
number density $\phi^*$ sets the normalization.
Both surveys use the same SDSS Petrosian photometry.

The HODs were evolved with redshift such that the target luminosity function is reproduced.
The target luminosity function can be used to calculate the magnitude threshold,
$\magr(n, z)$, as a function of redshift, which corresponds to a constant number
density $n$. For this sample with a fixed number density, the shape of the
HOD is kept constant (i.e. $\sigma_{\log M}$ and $\alpha$ are fixed), but
the mass parameters are all scaled by the same factor, so that the HOD produces the correct number density at each redshift.

\subsubsection{Unresolved haloes}

The MXXL simulation has a halo mass resolution limit of $\sim 10^{11}~\hMpc$.
There are some faint galaxies at low redshifts which are brighter than the magnitude threshold, 
but which are missing from the mock since they reside in haloes that fall below the simulation mass
limit. Therefore, there would be incompleteness
in the mock at low redshifts, which would be problematic for certain uses of 
the mock catalogue, e.g. testing
fibre assignment algorithms, and measuring the luminosity function. To ensure that the mock
is complete, haloes below the MXXL mass resolution were added to the lightcone.
Fits were done to the halo mass function in narrow redshift bins,
which were extrapolated to lower masses. Halo masses were drawn randomly from the extrapolated
mass function. Since there is no particle information available for the MXXL
simulation, the unresolved haloes were positioned randomly, so they are unclustered.

\subsubsection{Colour distributions}
\label{sec:colour_distributions}

Each galaxy is assigned a $\gr$ colour, following a parametrisation of the colour-magnitude
diagram from SDSS and GAMA. 
For the colours, both surveys use SDSS DR7 model magnitudes \citep{Abazajian2009}.
In a narrow redshift and absolute magnitude bin, the
colour distribution can be modelled as a double-Gaussian, with parameters
$\mu_\mathrm{red}(\magr)$, $\sigma_\mathrm{red}(\magr)$,
$\mu_\mathrm{blue}(\magr)$ and $\sigma_\mathrm{blue}(\magr)$ describing the
mean and width of the red and blue sequences, and $f_\mathrm{blue}(\magr)$ is the fraction
of blue galaxies, which gives the relative contribution of the two sequences. 
The parametrisation of \citet{Skibba2009} was used, where these are all
modelled as linear functions of magnitude,
but adjustments were made at the faint end to improve the agreement with measurements from
GAMA. Redshift evolution was also added to these parameters. 

This describes the total colour distribution, but central and satellite galaxies
have different probabilities of being red or blue. To model this, two
additional parameters were included: the fraction of galaxies that are satellites, $f_\mathrm{sat}(\magr)$,
and the mean colour of the satellite galaxies, $\mu_\mathrm{sat}(\magr)$.
$f_\mathrm{sat}(\magr)$ was measured from the mock, and the $\mu_\mathrm{sat}(\magr)$ from \citet{Skibba2009}
was used, with some evolution added.

To assign a colour to each galaxy, a random number is first generated to decide whether
a galaxy should fall on the red or blue sequence. This probability is different
for central and satellite galaxies. A random $\gr$ colour is then drawn from the appropriate
Gaussian distribution.

\subsubsection{$k$-corrections}

The final step of creating the mock is to convert the absolute magnitudes, $\magr$, into the
observed $r$-band apparent magnitude,
\begin{equation}
r = \magr + 5\log_{10}D_L(z)+ 25 + \kcorrr(z),
\end{equation}
where $D_L(z)$ is the luminosity distance (in $\hMpc$), and $\kcorrr(z)$ is the $k$-correction,
which takes into account the shift in the bandpass with redshift.

We use a set of $k$-corrections from the GAMA survey, where each galaxy has a 4th order polynomial 
$k$-correction,
\begin{equation}
\kcorrr(z) = \sum^4_{i=0} A_i (z-0.1)^{4-i}.
\end{equation}
The GAMA galaxy sample was split into 7 equal bins of $\gr$ colour, and the median $k$-correction was found
in each bin \citep[see figure~13 of][]{Smith2017}. The $k$-corrections were then interpolated between bins. A similar set of $k$-corrections
was measured from the data in the $g$-band, which are used when converting $g-r$ colours
from the rest frame to the observer frame. The observer-frame colour is 
\begin{equation}
g-r = {}^{0.1}(g-r) + {}^{0.1}k_g(z) - {}^{0.1}k_r(z).
\end{equation}

Finally, an apparent magnitude cut is applied to the galaxies in the mock.
In \citet{Smith2017}, a magnitude limit of $r<20$ was used, 
corresponding to the faint limit that was originally planned for the BGS faint sample. In this work, we extend this limit to $r<20.2$, to encompass the final BGS FAINT selection.

\section{Halo interpolation}
\label{sec:interpolation}

In this section, we investigate different halo interpolation schemes which are used
when building the halo lightcone. 

Firstly, we found that there was a bug in the original halo lightcone, where
the snapshot redshifts had been rounded to two decimal places. This led to variations
in the line-of-sight velocity dispersion, in narrow redshift bins. At the redshift of each snapshot,
the velocity dispersion in the lightcone agreed with the measurements in the snapshot. 
However, half way between snapshots, this rounding of the redshifts led to the velocities sometimes
being boosted, and other times reduced, by as much as a factor of $\sim 2$ in the 
most extreme case. This has been corrected in the new MXXL mock. 

To compare different halo interpolation methods, we create halo lightcones
using the interpolation methods described in the following section. Each
halo lightcone is then populated with galaxies, following the same HOD
methodology outlined in Section~\ref{sec:mock_creation_method}.

\subsection{Interpolation schemes}

\begin{figure} 
\centering
\includegraphics[width=\linewidth]{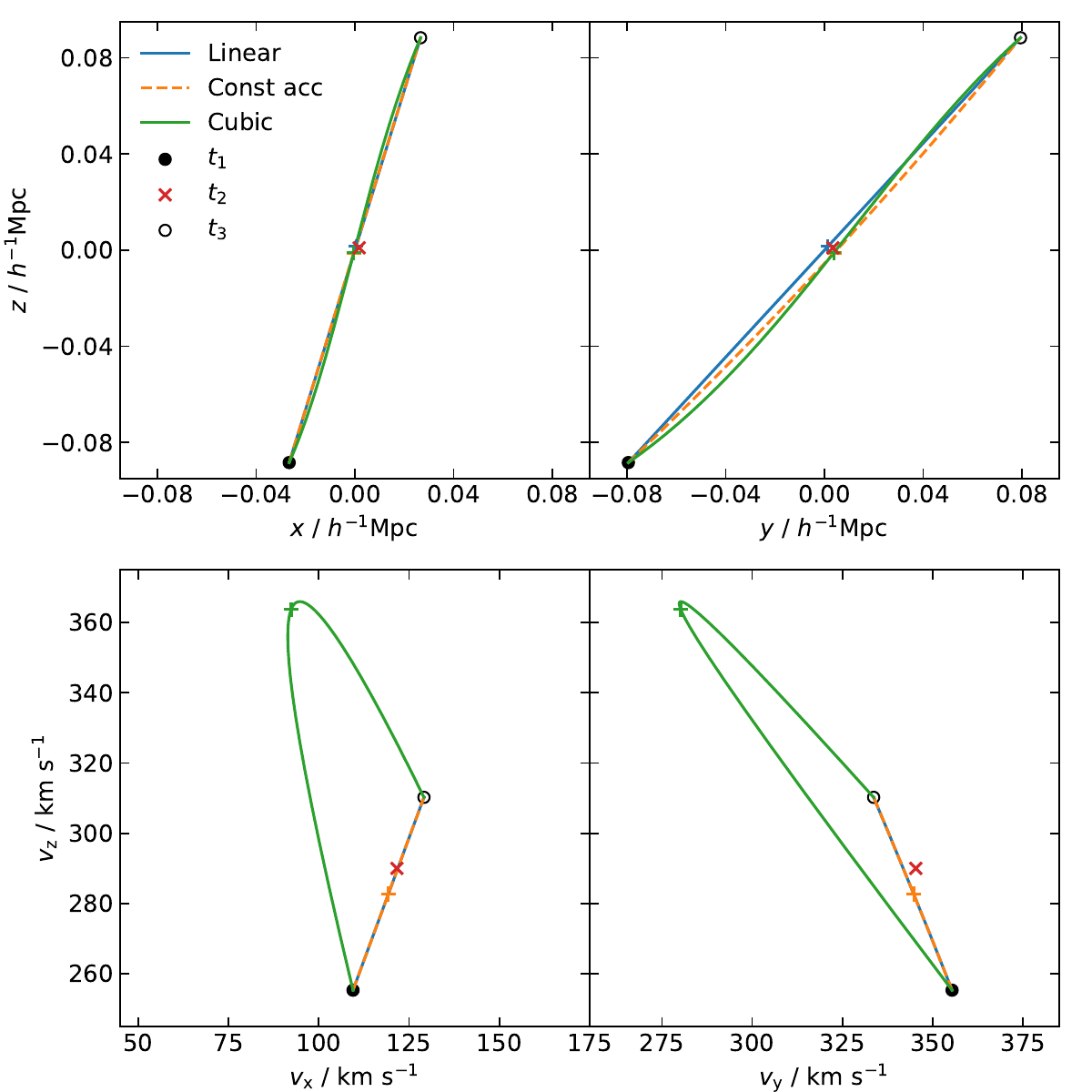}
\caption{\textit{Upper panels}: example showing the interpolated path of a halo, projected in two orthogonal planes. The
initial and final positions of the halo at $t_1$ and $t_3$ are shown by the black and open circles, respectively.
The curves show the interpolated path, using linear interpolation (blue), interpolation with a constant acceleration
(orange dashed), and cubic interpolation (green). The `true' position at the intermediate snapshot time, $t_2$, is shown by the red cross, while the interpolated positions at the same time are indicated
by the plus symbols.
\textit{Lower panels}: the same as above, but showing the
interpolated velocity vectors.}
\label{fig:halo_interpolation}
\end{figure}

Halo interpolation is done between two simulation snapshots at times $t=t_1$ and $t=t_2$, 
in units where $t_1=0$ and $t_2=1$.

\subsubsection{Cubic interpolation}
Each component of the position vector is interpolated following a cubic function,
while the velocities follow a quadratic function that is consistent with
the positions,
\begin{equation}
\begin{split}
x(t) &= A + Bt + Ct^2 + Dt^3 \\
v(t) &= B + 2Ct + 3Dt^2,
\end{split}
\end{equation}
where the coefficients are determined by the requirements
that $x(t_1) = x_1$, $v(t_1) = v_1$, $x(t_2) = x_2$ and 
$v(t_2) = v_2$. This gives
$A = x_1$ and $B = v_1$, which are the initial positions and velocities, and the other terms are
\begin{equation}
\begin{split}
C &= -2v_1 - 3x_1 - v_2 + 3x_2 \\
D &= v_1+v_2 + 2(x_1-x_2).
\end{split}
\end{equation}

\subsubsection{Linear interpolation}

In this scheme, velocities are interpolated linearly, assuming a constant acceleration,
and positions are also interpolated linearly, assuming a constant velocity,
\begin{equation}
\begin{split}
x(t) &= x_1 + \overline{v} t \\
v(t) &= v_1 + \overline{a} t,
\end{split}
\end{equation}
where $\overline{v} = x_2 - x_1$ is the average velocity in the $x$-direction (since we are working in units where $t_2-t_1=1$), and $\overline{a} = v_2 - v_1$ is
the average $x$-component of the acceleration. Similar equations are used in the $y$ and $z$-directions. Since the positions and velocities are interpolated independently 
of each other, the rate of change of the positions will not be consistent with the velocities.

\subsubsection{Constant acceleration}

This is similar to linear interpolation, where velocities are interpolated linearly, assuming a constant
acceleration. However, the interpolated positions are consistent with the constant acceleration,
\begin{equation}
\begin{split}
x(t) &= \overline{x} + \overline{v}\left(t-\frac{1}{2}\right) + \frac{\overline{a}}{2}\left(t-\frac{1}{2}\right)^2 - \frac{\overline{a}}{8} \\
v(t) &= v_1 + \overline{a}t,
\end{split}
\end{equation}
where $\overline{v}$ and $\overline{a}$ are the average velocity and acceleration, and $\overline{x}$ is the
mean position, $\overline{x} = (x_1+x_2)/2$. Even though the positions and velocities are consistent with
each other, it is not guaranteed the the initial and final positions will be correct. The extra term
$-\overline{a}/8$ ensures that at $x(t_1)=x_1$ and $x(t_2)=x_2$.

\subsubsection{No interpolation}

We also consider the case where no interpolation is applied to the halo catalogue, creating a mock
from a single snapshot at $z=0.14$. This snapshot is chosen as it is at the median redshift of the
galaxy sample used for assessing the two-point clustering statistics in Section~\ref{sec:interpolation_clustering}. Galaxy positions in the cubic box are
converted to equatorial coordinates, with the observer in the same location as the halo lightcone.
To ensure that the mock has the same
evolving luminosity function as the other lightcones, a rescaling is applied to the magnitudes. This rescaling preserves the ranking of the $r$-band magnitudes at each redshift.

\subsection{Comparing interpolation schemes}

\begin{figure} 
\centering
\includegraphics[width=\linewidth]{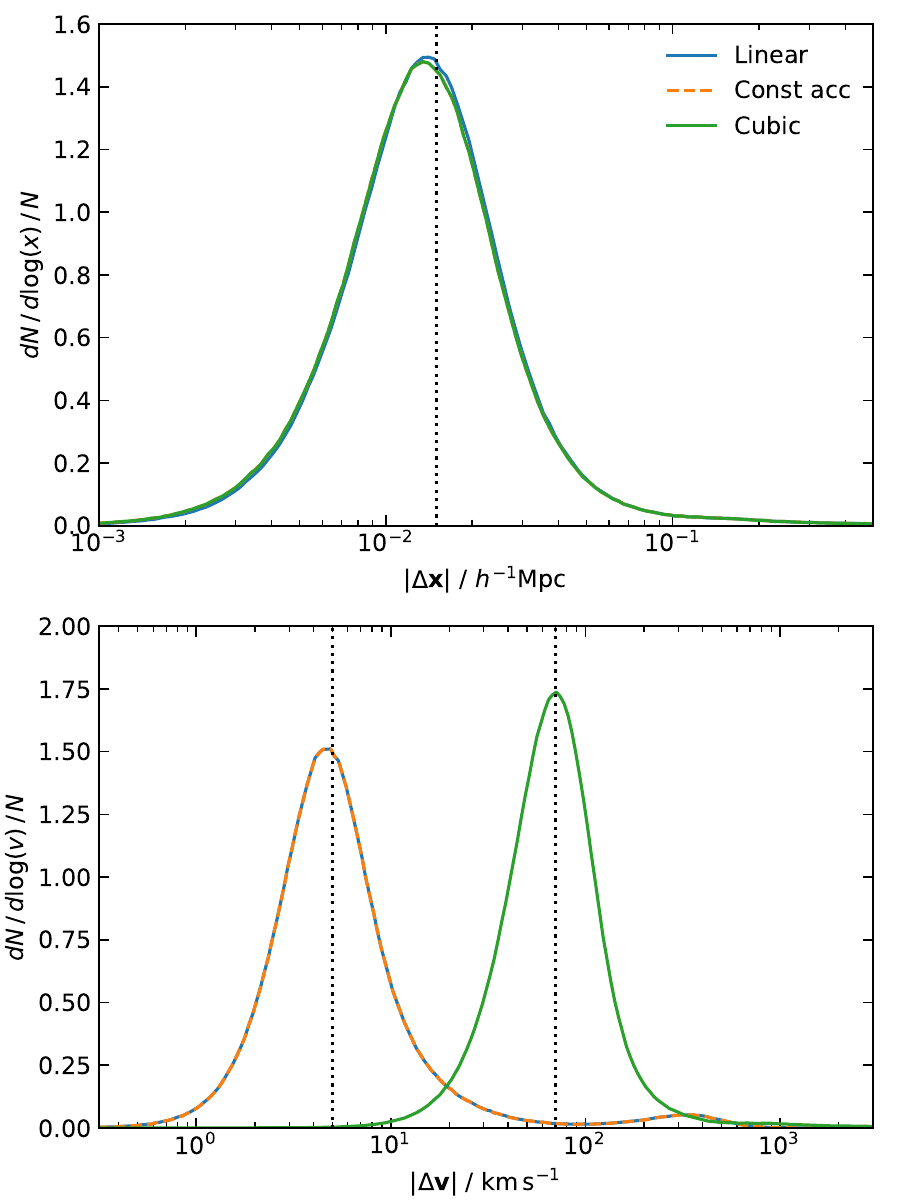}
\caption{\textit{Top panel}: distribution of $|\Delta \textbf{x}| = |\textbf{x}_{t_2}^\mathrm{interp} - \textbf{x}_{t_2}^\mathrm{true}|$, which is
the difference between the true and interpolated positions at the intermediate snapshot time, $t_2$. Haloes are interpolated between the snapshots $z_1=0.144$ and $z_3=0.089$, where the intermediate
snapshot is $z_2=0.116$. Linear
interpolation, interpolation with a constant acceleration, and cubic interpolation are shown in blue, orange and green, respectively.
\textit{Bottom panel}: as above, but showing the distribution of velocity offsets, 
$|\Delta \textbf{v}| = |\textbf{v}_{t_2}^\mathrm{interp} - \textbf{v}_{t_2}^\mathrm{true}|$}
\label{fig:interpolation_differences}
\end{figure}

\begin{figure} 
\centering
\includegraphics[width=\linewidth]{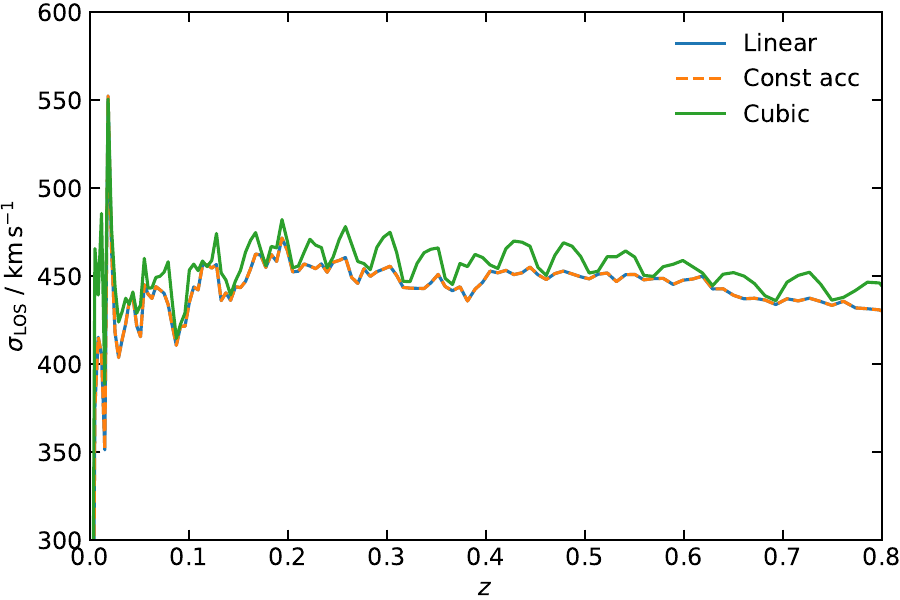}
\caption{Line-of-sight velocity dispersion, $\sigma_\mathrm{LOS}$, in the halo lightcone, as a function of redshift. This is measured for
the different interpolation schemes in narrow redshift bins, which we define as the difference between the 84th and 16th percentiles of the
velocity distribution. Linear interpolation is in blue, constant acceleration interpolation is the orange dashed curve,
and cubic interpolation is in green.}
\label{fig:velocity_dispersion}
\end{figure}

An example of halo interpolation is shown in Fig.~\ref{fig:halo_interpolation}, comparing the
positions and velocities produced by the different interpolation schemes, for
a halo interpolated between the snapshots at $z_1=0.144$ and $z_3=0.089$ with mass $\sim 2.5 \times 10^{12}~\hMsun$. Here, we have skipped the intermediate snapshot, $z_2=0.116$, in order to compare the interpolated positions and velocities 
with the `true' values measured in the snapshot.
In the upper panels, the position trajectories are similar
for the different schemes. 
Small differences can be seen between the different curves in the top right panel, but
the interpolated positions at $t_2$ are all very close to the position measured at the central snapshot.
The lower panels compare the interpolated velocities, and here the differences are much more apparent.
By construction, the linear and constant acceleration cases are identical, since they both use the 
same constant acceleration. However, for the case of cubic interpolation, we see large 
deviations in the velocity. 
At time $t_2$, the velocity of the halo measured in the central snapshot is roughly half way 
between the initial and final velocities, $v_1$ and $v_3$, which is what we expect for most haloes.
Linear and constant acceleration interpolation produce velocities close to this, while for
cubic interpolation, each velocity component can be either much larger or much smaller.
In this example, the $z$-component of velocity increases by $\sim 40\%$.
This is a typical example, and large variations like this are seen for the majority of haloes,
covering a wide range of halo mass. Note that since we have skipped the middle snapshot, 
the differences seen here are larger than in the final lightcone, which uses all snapshots.
Linear and constant acceleration interpolation are in close agreement for most haloes, 
but there are cases where the positions differ (e.g. if the velocity in one direction changes sign).

In Fig.~\ref{fig:halo_interpolation} we compared the interpolated positions
of a single halo to its true position at the intermediate snapshot.
The upper panel of Fig.~\ref{fig:interpolation_differences} shows the distribution of the difference
in position, $|\Delta \textbf{x}| = |\textbf{x}_{t_2}^\mathrm{interp} - \textbf{x}_{t_2}^\mathrm{true}|$, for the different interpolation methods.
This distribution is measured for a subset of 4 million haloes, covering the full range of halo masses,
at the same intermediate redshift $z_2=0.116$. The three different interpolation methods
produce an almost identical distribution, which peaks at $|\Delta \textbf{x}| \sim 15~\hkpc$ 
with only $\sim 2\%$ of haloes with $|\Delta \textbf{x}| > 0.1~\hMpc$. 
The lower panel of Fig.~\ref{fig:interpolation_differences} shows the difference in 
velocities for the same set of interpolated haloes, $|\Delta \textbf{v}| = |\textbf{v}_{t_2}^\mathrm{interp} - \textbf{v}_{t_2}^\mathrm{true}|$. 
For linear interpolation, this distribution peaks at $\sim 5~\kms$, while the
velocity differences are much larger for the case of cubic interpolation, which
peaks at $\sim 70~\kms$. This is consistent with the halo in Fig.~\ref{fig:halo_interpolation},
and shows that it is a typical example of a halo. The distribution of velocities with linear
interpolation is also bimodal, with a second smaller peak at $\sim 400~\kms$. Linear interpolation
performs poorly for haloes which undergo mergers, and the haloes in the secondary peak have a very large difference between their initial and final mass at $t_1$ and $t_3$. The bimodality is not seen in the case of cubic interpolation, but halo mergers are responsible for the tail which extends beyond $1000~\kms$. 

The line-of-sight velocity dispersion is shown in Fig.~\ref{fig:velocity_dispersion},
for haloes in the lightcone in narrow redshift bins, with $M_\mathrm{h} > 10^{11}~\hMsun$, which is measured from
the difference between the 84th and 16th percentiles of the velocity distribution. 
Since the linear
and constant acceleration interpolation schemes both have the same velocities, the
curves are identical, and the velocity dispersion is fairly flat as a function of redshift.
However, for cubic interpolation, the velocity dispersion increases half way between snapshots,
where the interpolated velocity can be very different from the initial and final velocities,
before dropping down again at the redshift of the next snapshot.

In the simulation, particles travel along smooth trajectories, so it is surprising
that cubic interpolation poorly describes the motion of the haloes. The halo finder is run
independently at each snapshot, which means that there are differences in the set
of particles that are identified as belonging to the halo at each simulation output. 
Between snapshots, the halo will accrete particles and undergo halo mergers.
This can lead to positions and velocities at two snapshots that are inconsistent with
each other.

When creating mocks, we want to use an interpolation scheme which produces
galaxy clustering that evolves smoothly with redshift, and avoids adding
systematics due to the inconsistencies between positions and velocities.
In the next section, we compare the two-point clustering statistics of
mocks created using these different halo interpolation schemes.

\subsection{Galaxy clustering}
\label{sec:interpolation_clustering}

\begin{figure*} 
\centering
\includegraphics[width=0.48\linewidth]{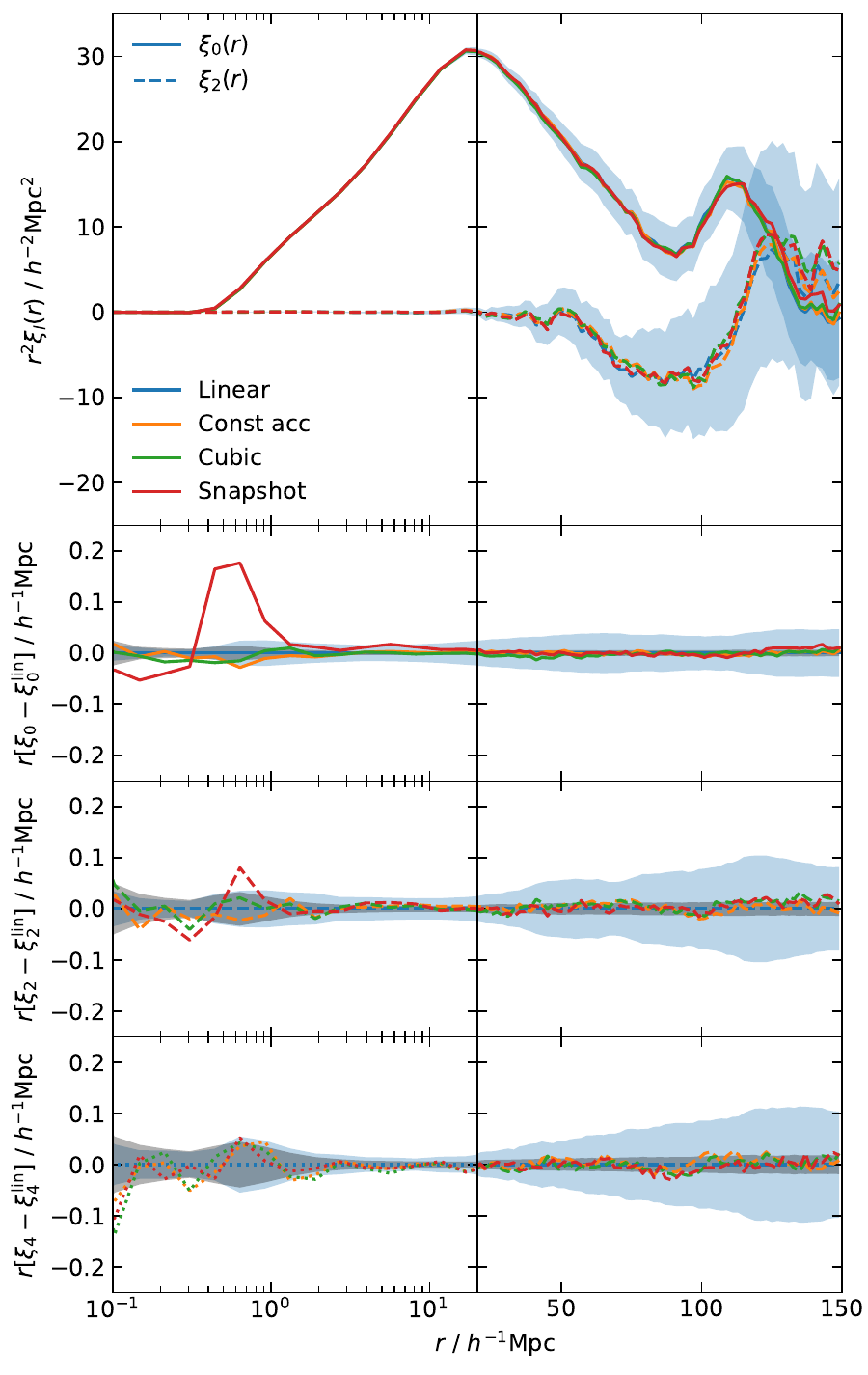}
\includegraphics[width=0.48\linewidth]{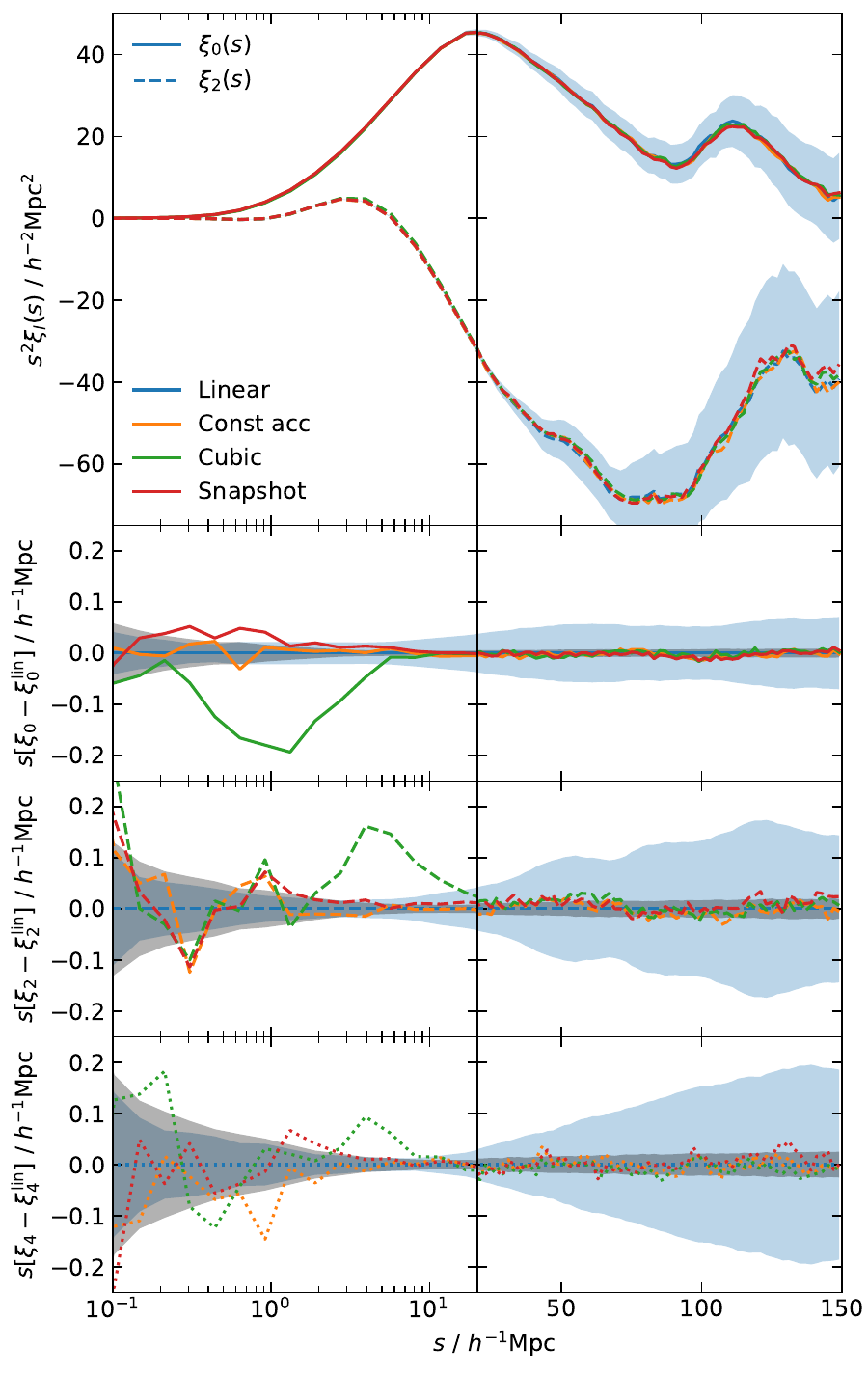}
\caption{\textit{Left}: Correlation function multipoles in real space, for a sample of central galaxies with magnitude
$\magr < -20$ and redshift $z<0.2$. The top panel shows the monopole (solid lines), and quadrupole (dashed), with a linear $x$-scale on large scales, and logarithmic scale for $r<20$. 
The different coloured lines are for mocks with linear interpolation (blue), constant acceleration interpolation (orange),
cubic interpolation (green), and a mock with no interpolation, built from a single snapshot (red). The second panel show the differences 
in the monopole measurements, scaled by $r$, relative to linear interpolation. The third and fourth panels are the same, but for the quadrupole and hexadecapole, respectively. The 
blue shaded region is the jackknife error, using 100 jackknife samples, and the grey shaded region is a jackknife error in $r\Delta \xi$, 
averaged over all pairs of mocks. \textit{Right}: Correlation function multipoles in redshift space.}
\label{fig:xi_interpolation}
\end{figure*}

The two-point correlation function of galaxies in the lightcone mocks, with
different halo interpolation schemes, are shown in Fig.~\ref{fig:xi_interpolation}, where the mock catalogue is cut to a volume limited sample of central galaxies, 
with $\magr < -20$ and $z<0.2$. We do not consider satellite galaxies, since
they are positioned randomly around each central galaxy and are not affected by the halo interpolation
method. We measure the 2D correlation function, $\xi(s,\mu)$, where $\mu$ is the cosine of the angle between the line-of-sight and pair separation vector. This is then decomposed into Legendre multipoles, where the monopole, $\xi_0(s)$, and quadrupole, $\xi_2(s)$ are non-zero in linear theory.

The left panel of Fig.~\ref{fig:xi_interpolation} shows the clustering
in real space.
The monopole is shown by the solid lines in the upper panel,
for the cases of linear interpolation, constant acceleration interpolation, cubic
interpolation, and no interpolation (i.e. a single snapshot). In real space, the monopole, $\xi_0(r)$, is equivalent to the real-space correlation function, $\xi(r)$.
The quadrupole is also plotted, which should be zero in real space, but on large scales we see some variation due to cosmic variance. Since the mocks were all constructed from the 
same simulation, with observer placed at the same position, the same
variations are seen on large scales.

The lower panels show the difference in the monopole, quadrupole and hexadecapole,
relative to
the linear interpolation case, scaled by a factor of $r$. The blue shaded
region indicates a jackknife error, using 100 jackknife samples, which is an
estimate of the error due to cosmic variance. On large scales, this error is 
larger than the scatter between curves, since they all have the same large-scale
structure. The grey shaded region shows the jackknife error on $r\Delta \xi$
\citep[equation 14 of][]{Grove2021}, which is an estimate of the noise between
pairs of simulations that have the same density field. This noise is measured for
each pair of mocks, and the average noise is plotted. On small scales, these two errors 
are comparable.

For these 4 different mocks, the real-space clustering on large scales ($r>20~\hMpc$) is in
good agreement with each other. On smaller scales, the different interpolation methods
remain in good agreement, but there is a large difference at $\sim 0.6~\hMpc$ compared
to the mock built from a single snapshot. It is likely
that these differences are because of how halo mergers are implemented in the 
lightcone, which is not perfect. 
In our scheme, each halo is interpolated towards the position of its descendant at the
next snapshot, so two merging haloes will be interpolated directly towards each other.
In reality, the merging haloes would not collide head-on, but the smaller halo would be
stripped as it passes close to the larger halo. When they merger, the pair separation 
of haloes in the snapshot will be larger compared to using interpolation, 
leading to a stronger clustering signal at $\sim 0.6~\hMpc$.
On even smaller scales there is a halo exclusion effect in the snapshot,
where if two haloes were positioned very closely together, the halo
finder would identify a single halo.
This results in the monopole being $\xi_0(r) \sim -1$ in the snapshot for 
$r \lesssim 0.3~\hMpc$. 
There is no constraint on the minimum separation when doing interpolation, 
so the clustering on these scales is stronger.

The right panel of Fig.~\ref{fig:xi_interpolation} shows the clustering in redshift space. 
When the effect of velocities are included, we still see good agreement between the different 
mocks on large scales. On small scales, the constant acceleration interpolation is in good
agreement with linear interpolation. Since both have the same velocities, and the 
real-space clustering is in good agreement, it is unsurprising that they also 
show good agreement in redshift space. However, there are some differences
compared to the other mocks. The mock built from a single snapshot still shows some excess
clustering on small scales in the monopole, but this is greatly reduced in redshift space
compared to in real space. At $0.6~\hMpc$, this small offset corresponds to $\sim 3\%$ in
$\xi_0$. Note that this plot is for central galaxies only. There are many pairs of satellite
galaxies on these small scales, which will reduce the difference further. However,
for the cubic interpolation case, we see much greater differences. There is a large
deficit in clustering in the monopole at $\sim 1~\hMpc$, and an excess in the 
quadrupole (and hexadecapole) at $\sim 5~\hMpc$. While the clustering in real space looks reasonable,
there are very large deviations in the interpolated velocities (which can be seen in the
example in Fig.~\ref{fig:halo_interpolation}), leading to offsets in the redshift-space 
clustering.

For all mocks, the clustering on large scales ($s\gtrsim 20~\hMpc$), which are used in a typical
RSD or BAO analysis are in good agreement. However, we conclude that it is better to use either 
linear interpolation or constant acceleration interpolation in the lightcone to improve the 
small-scale clustering measurements. We decide to adopt linear interpolation, since the 
equations describing the interpolation are very simple. Differences between this and the
single snapshot on small scales shows that improvements could be made to how halo mergers
are dealt with when building the lightcone. However in redshift space, these differences are very 
small, particularly when satellite galaxies are also included.

\section{Colour distributions}
\label{sec:colours}

\subsection{Colour assignment}

\begin{figure} 
\centering
\includegraphics[width=\linewidth]{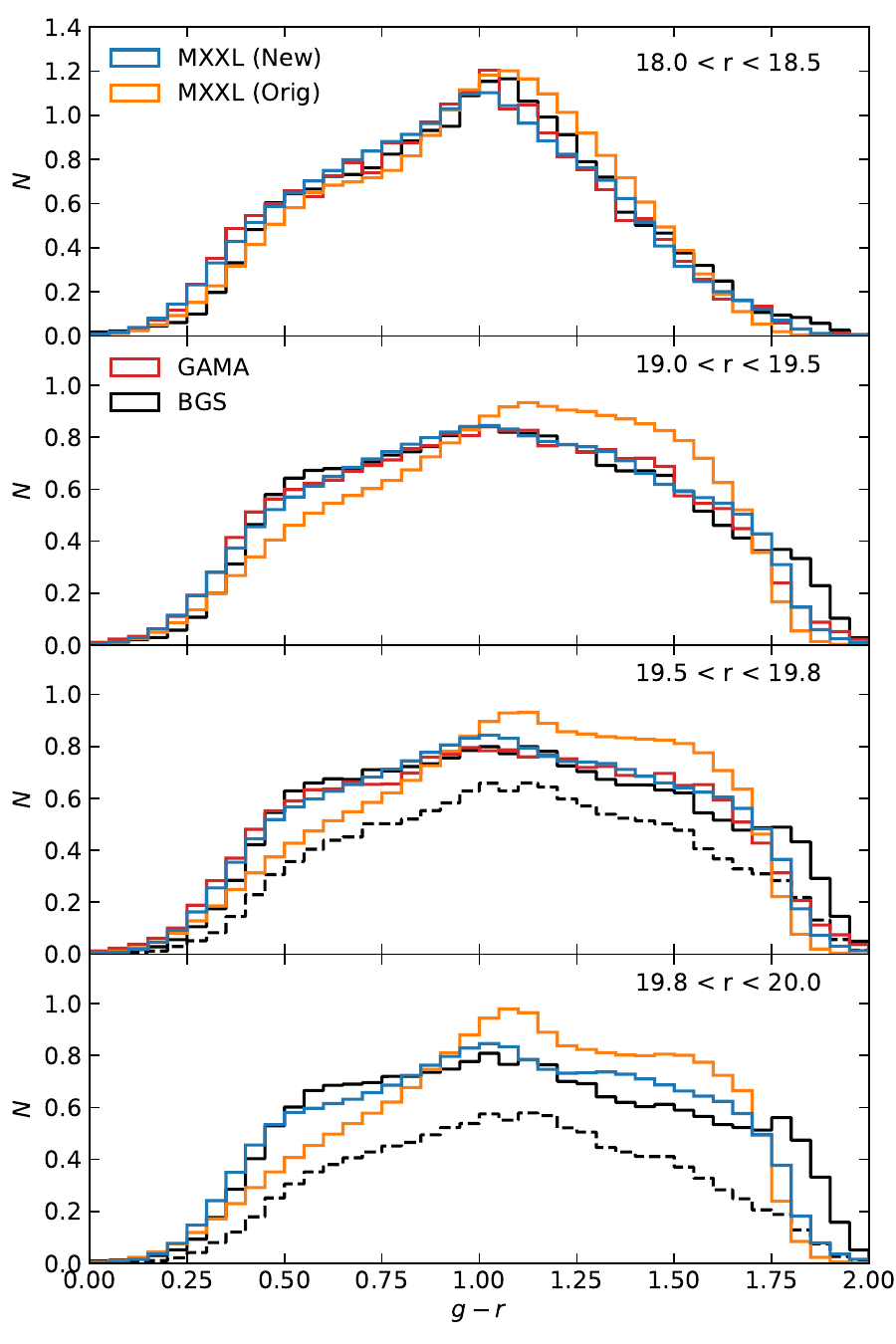}
\caption{Distribution of observer-frame $g-r$ colours, in bins of apparent
$r$-band magnitude. The magnitude bins decrease in brightness from the top
to bottom panel. The new MXXL mock is shown by the blue histogram, while the
original MXXL mock is in orange, and GAMA data is in red. The colour distributions
from the DECaLS BGS targets are shown in black, with (dashed) and without (solid) the BGS colour selection
of the BGS FAINT galaxies (for $r>19.5$).
More details regarding the magnitude definitions used for these curves are provided in
Section~\ref{sec:improved_colours}.}
\label{fig:observed_colour_distribution}
\end{figure}

\begin{figure} 
\centering
\includegraphics[width=\linewidth]{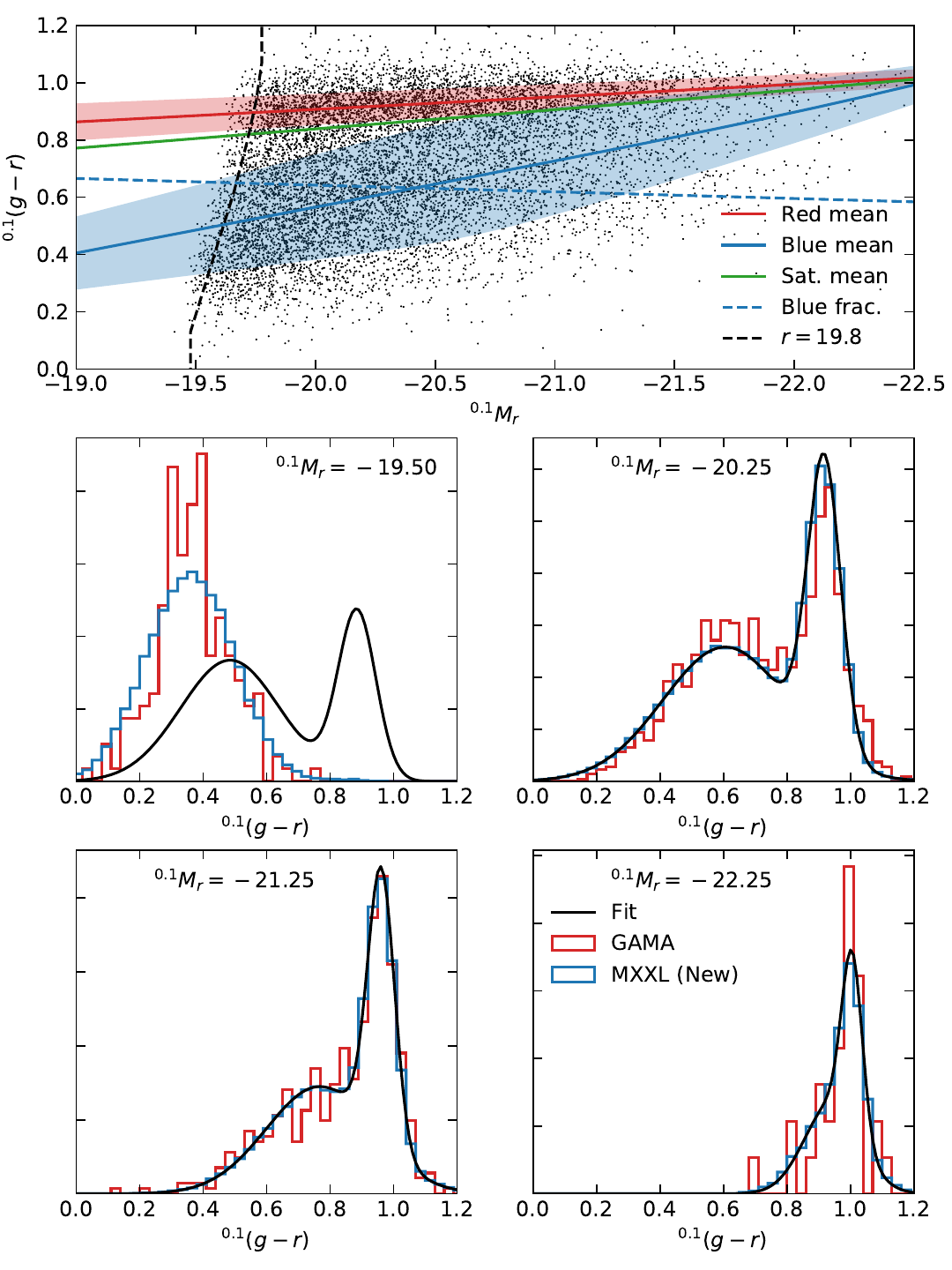}
\caption{\textit{Top panel}: Colour-magnitude diagram from GAMA, in a narrow 
redshift slice at $z=0.225$, where colours and magnitudes are in the
rest-frame, $k$-corrected to $z=0.1$ (black points). Red and blue solid lines show the fits
to the red and blue sequence, respectively, where the shaded regions indicate
1$\sigma$. The green line is the mean satellite colour. The blue dotted line shows the
fraction of blue galaxies as function of absolute magnitude (for this line only the y-axis is a fraction and not a colour). 
The black dotted line shows how, as a function of colour, the $r=19.8$ apparent magnitude limit maps to absolute magnitude at $z=0.225$.
\textit{Lower panels}: distribution of colours in narrow absolute
magnitude bins, for GAMA (red), the fit (black) and the new MXXL mock (blue). The magnitude bins have width 0.2, and are centred on the values indicated in each panel.}
\label{fig:colour_distribution_fits}
\end{figure}

Colours are assigned to the MXXL mock catalogue using a parametrisation
of the GAMA colour-magnitude diagram. As described in
Section~\ref{sec:colour_distributions}, the bimodal rest-frame
$\gr$ colour distribution in bins of absolute magnitude and redshift
is well modelled by a double-Gaussian function. The 5 parameters are
$\mu_\mathrm{red}$, $\sigma_\mathrm{red}$, $\mu_\mathrm{blue}$, $\sigma_\mathrm{blue}$, $f_\mathrm{blue}$, which are the mean and rms
of the red and blue sequences, and the total fraction of galaxies
that are blue. In addition, the mean satellite colour and satellite
fraction, $\mu_\mathrm{sat}$ and $f_\mathrm{sat}$, are also
required to model the differences in colour between centrals and satellites \citep[see][]{Skibba2009}.


In the original MXXL mock, the colours and satellite fraction are linear functions of absolute magnitude,
which were fit to the colour distributions measured from SDSS \citep{Skibba2009}, but modified to bring
the colour distributions at the faint end into better agreement with GAMA. These were then evolved with redshift,
keeping the shape fixed, but modifying the amplitude, so that the colours matched the distributions from
the GAMA survey \citep{Smith2017}. As shown in figure~14 of \citet{Smith2017}, the rest-frame
colour distributions are in reasonable agreement with GAMA over a wide range of redshifts.

However, while the rest-frame $\gr$ colour distributions look reasonable, there are discrepancies in the
observer-frame colours compared to GAMA. This is shown in Fig.~\ref{fig:observed_colour_distribution}, in several
bins of $r$-band apparent magnitude. The orange histogram is the $g-r$ distribution measured in the MXXL
mock with the original colour distributions, while the red histogram shows the measurements from the 
GAMA survey. The discrepancies between them are greater at faint magnitudes, where there is an excess of
red galaxies, and deficit of blue galaxies in the mock, compared to the data. While the rest-frame
colours are in reasonable agreement with GAMA, there are differences at faint magnitudes, which are seen much
more prominently in the observer-frame colours. The other histograms in Fig.~\ref{fig:observed_colour_distribution} are discussed at the end of Section~
\ref{sec:improved_colours}.

\subsection{Improving the colour distributions}
\label{sec:improved_colours}

To improve the colour distributions in the mock, we modify the functional form of the parameters
$\mu_\mathrm{red}(\magr)$, $\sigma_\mathrm{red}(\magr)$, $\mu_\mathrm{blue}(\magr)$,
$\sigma_\mathrm{blue}(\magr)$ and $f_\mathrm{blue}(\magr)$ to use a broken linear function, with a smooth transition at the break.
These fits are all shown in the upper panel of Fig.~\ref{fig:colour_distribution_fits}, at $z=0.225$.
We find that this provides a better fit to the data, compared to linear functions. This was then fit to
the colour distributions of GAMA, in bins of redshift of width $\Delta z = 0.05$, for magnitudes brighter
than the $r=19.8$ magnitude cut of GAMA. 
However, these fits assume that the galaxy sample in GAMA is complete, and
do not take into account incompleteness at the $r=19.8$ limit.
To check this, the fits in each redshift bin were used to assign galaxy colours in the
MXXL mock in the same redshift bin, and the $r=19.8$ cut was applied. This allowed us to compare
the observer-frame colour distributions at the faint limit, and adjust the fits so
that the MXXL mock was in agreement with the GAMA data. To create the final
MXXL mock, colours were assigned using the fits at all redshift, interpolating
between them. For the mean satellite colour, we set it to be between the
red and blue sequences using
\begin{equation}
\mu_\mathrm{sat}(\magr) = f \mu_\mathrm{red}(\magr) + (1-f) \mu_\mathrm{blue}(\magr),
\end{equation}
and find that using $f=0.1(\magr-20.5)+0.8$, with $0<f<1$, gives reasonable colour-dependent clustering in the mock (see Section~\ref{sec:colour_dependent_clustering}).


To make sure that the $k$-corrections are consistent between the mock and the GAMA
data, the same colour-dependent $k$-corrections are applied to the data to
calculate absolute $r$-band magnitudes and rest-frame $\gr$ colours.
Since the $k$-correction depends on the rest-frame colour, a root finding
procedure is done to find the rest-frame colour that produces self-consistent $k$-corrections.


The colour-magnitude diagram of GAMA is shown by the black points in the upper panel of Fig.~\ref{fig:colour_distribution_fits}
along with the fits, for galaxies at $z=0.225$ in a redshift bin of width 0.02. The black dashed line 
indicates the magnitude cut corresponding to the $r=19.8$ limit of GAMA at $z=0.225$. 
At high redshifts, the $r=19.8$ cut corresponds to a brighter absolute magnitude for
red galaxies than for blue galaxies, due to the colour-dependent $k$-corrections used. 
Some galaxies appear to be fainter than this limit due to the 
width of the redshift bin plotted. The lower panels show histograms of the same $\gr$ colour
distribution, in bins of absolute magnitude of width 0.2, centred on the values indicated
in each panel. The distributions from
GAMA are shown by the red histograms, while the black curves are the fits. The blue histogram
shows the result of using the colour distributions to assign colours to galaxies in MXXL, with a 
$r=19.8$ magnitude cut applied to match the GAMA data. For the bright magnitude bins, the double-Gaussian
fits are in very good agreement with the GAMA data, however the faintest magnitude bin is affected by incompleteness
at the $r=19.8$ limit. After assigning colours to the mock and applying a $r=19.8$ cut, the distribution of the remaining galaxies is in good agreement with GAMA. This figure only shows the colours in one
redshift bin, but our fits show similar good agreement with GAMA over the full redshift range of the mock. 

The observer-frame colour distributions in the mock, with the updated colour distributions, are shown by the blue histograms in
in Fig.~\ref{fig:observed_colour_distribution}. 
The MXXL colour distributions are in good agreement with GAMA (in red) down to the faint
limit of $r=19.8$. We also plot the colour distribution of targets from the BGS. These photometric BGS targets are from DECaLS, which is part of the DESI Legacy Imaging Survey, and these targets have been matched to the galaxies in GAMA. The imaging was done using the Dark Energy Camera \citep[DECam;][]{Flaugher2015}, which has slightly different passbands than GAMA. We do not correct for this, but the correction is small \citep[][]{Zarrouk2021}.\footnote{Note that GAMA uses Petrosian $r$-band magnitudes and SDSS DR7 model magnitudes for the $g-r$ colours \citep{Abazajian2009}. The DESI BGS uses Tractor model magnitudes \citep{Lang2016,Dey2019}.} 
Black
solid histograms are for all BGS targets within each apparent magnitude bin, selected only on apparent magnitude. The black
dashed histograms show the subset of BGS targets after also applying the BGS FAINT selection, which depends on colour and fibre magnitude. We find that the colours of the BGS targets are very similar to GAMA. When we extrapolate our fits to magnitudes fainter than $r=19.8$, the colour distributions remain in good agreement
with the BGS targets, and are greatly improved compared to the original MXXL mock catalogue.

In the bottom panel, the bump at red colours
is due to stellar contamination in the sample. This can be seen in figure~7 of \citet{Ruiz2021}, where there is a vertical spur of objects with very red $r-z$ colours.

\subsection{Colour-dependent clustering}
\label{sec:colour_dependent_clustering}

\begin{figure*} 
\centering
\includegraphics[width=\linewidth]{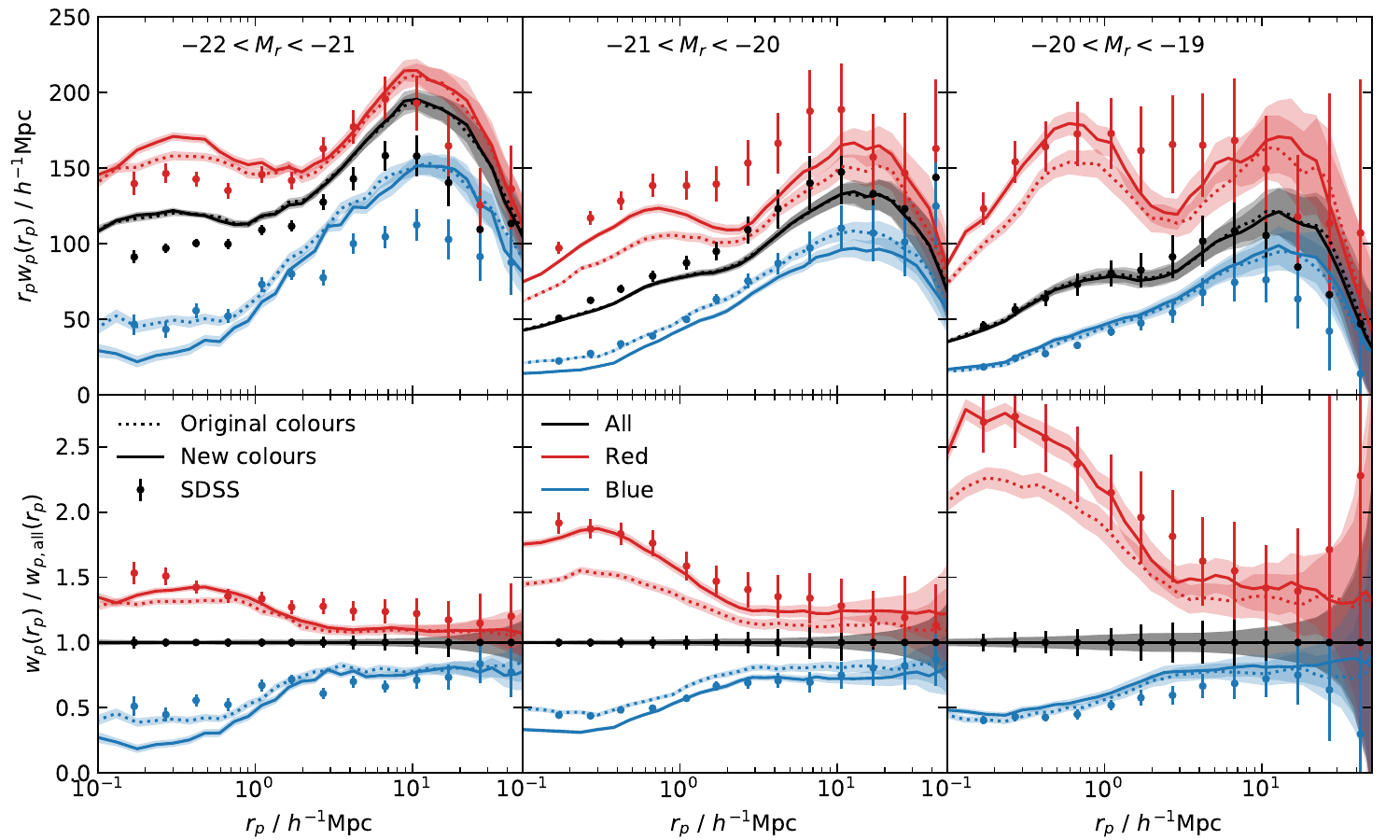}
\caption{Colour-dependent clustering in the MXXL mock at low redshifts, compared to measurements from
SDSS. The top panels show the projected correlation function, $w_p(r_p)$, for 3 magnitude bin samples of
$-22 < \magr < -21$ (left), $-21 < \magr < -20$ (centre), $-20 < \magr < -19$ (right). Points with error
bars are the SDSS measurements from \citet{Zehavi2011}, solid lines are from the MXXL mock with the
new colour distributions, and dotted lines are from the mock with the original colour distributions.
The shaded regions are jackknife errors, using 100 jackknife samples. All galaxies in each sample are
shown in black, while the split into red and blue galaxies is shown in red and blue respectively,
with a colour cut $\gr_\mathrm{cut}=0.21 - 0.03\magr$. The lower panels show the ratio of the
projected correlation functions, relative to the clustering of all galaxies, highlighting
the colour dependence of the clustering.}
\label{fig:wp_colour_sdss}
\end{figure*}

\begin{figure} 
\centering
\includegraphics[width=\linewidth]{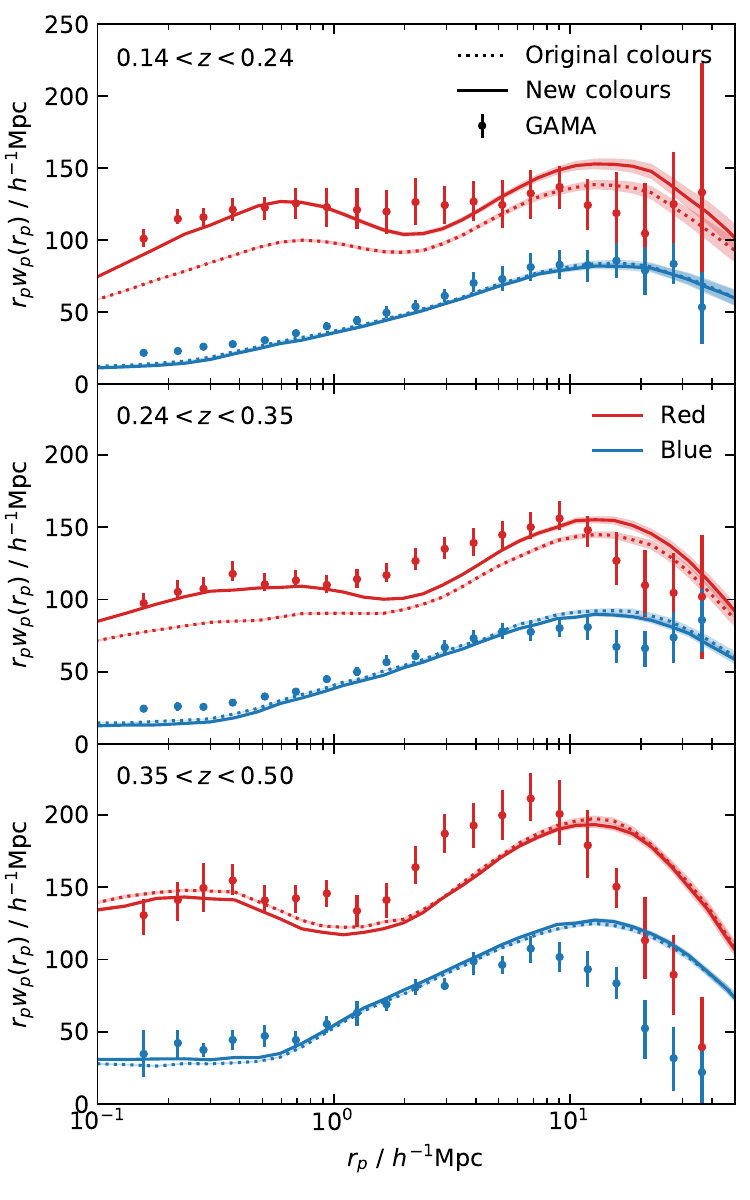}
\caption{Colour-dependent clustering in the MXXL mock in several redshift bins, compared to
measurements from GAMA. Each panel shows the projected correlation function, $w_p(r_p)$,
for a different redshift range, increasing in redshift from top to bottom. 
Points with error bars show the GAMA measurements of \citet{Farrow2015}, split into red
and blue galaxies. The measurements from the mock with the new colour distributions is 
shown by the solid lines, with errors calculated from 100 jackknife regions, and dotted
lines are from the mock using the original colours. The same colour cut is applied to the
mock as the GAMA data.}
\label{fig:wp_colour_gama}
\end{figure}

\begin{figure} 
\centering
\includegraphics[width=\linewidth]{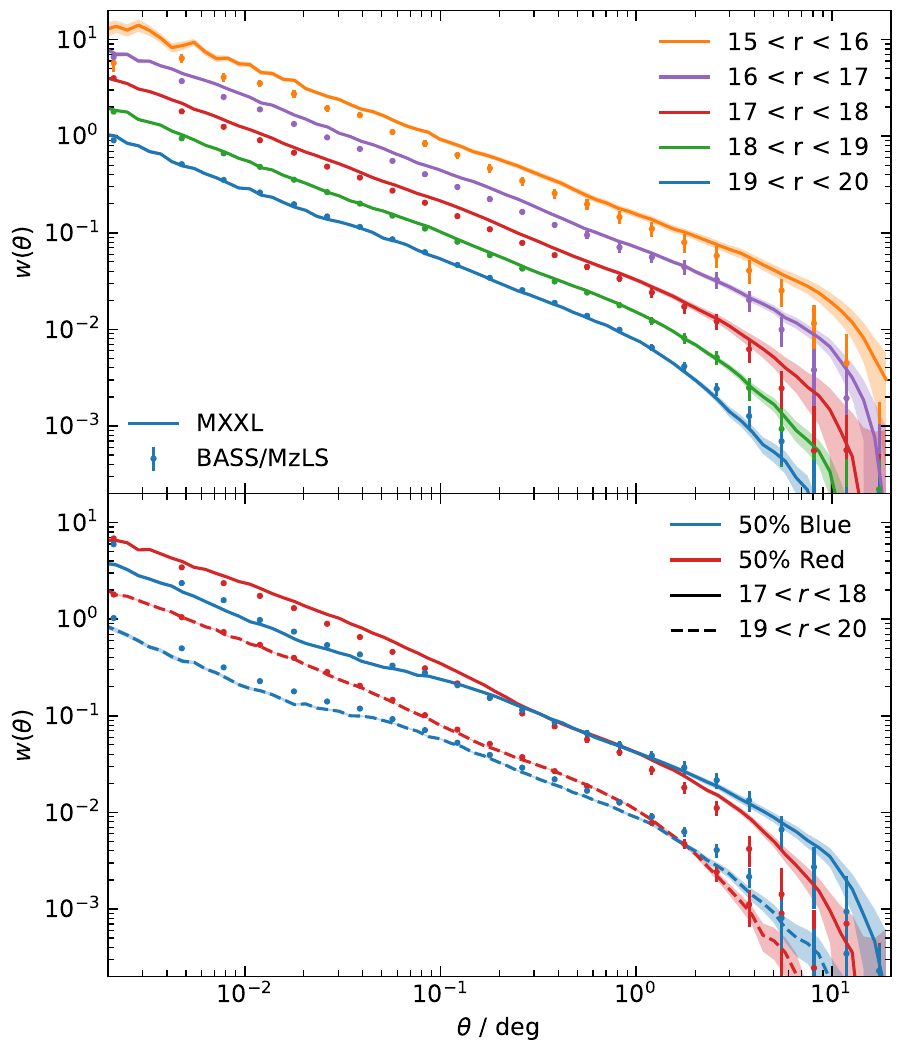}
\caption{\textit{Top panel}: Angular correlation function, $w(\theta)$, in the MXXL mock (lines),
compared to measurements from BASS/MzLS \citep[points with error bars;][]{Zarrouk2021}, for different
$r$-band apparent magnitude bins, as indicated in the legend. Jackknife errors in the mock are calculated
using 100 jackknife regions. \textit{Bottom panel}: Angular clustering in two magnitude bins, as above,
but split into red and blue galaxies. The colour cut applied is based on median colour in each magnitude bin.}
\label{fig:wtheta}
\end{figure}

In this section, we assess the two-point clustering statistics of red and blue galaxies in the MXXL mock, compared to measurements from SDSS, GAMA and the BGS targets.

The projected correlation function, $w_p(r_p)$, is defined as
\begin{equation}
w_p(r_p) = 2 \int_0^{\pi_\mathrm{max}} \xi(r_p,\pi)d\pi,
\end{equation}
where $r_p$ and $\pi$ are the pair separations perpendicular and parallel to the line of sight, respectively. Integrating the 2-dimensional correlation function along the line of sight removes the effect of redshift space distortions. Since the measurements become noisy on large scales, the correlation function is integrated to a maximum scale $\pi_\mathrm{max}$.

The projected correlation function is shown in Fig.~\ref{fig:wp_colour_sdss} at low redshifts, compared to
measurements from SDSS \citep{Zehavi2011}. The upper panels show the projected correlation function in several 
bins of absolute magnitude, which are the same bins defined in table~1 of \citet{Zehavi2011}. Each sample is
also split into red and blue galaxies using the same $\gr_\mathrm{cut} = 0.21 - 0.03 \magr$.
To be consistent with the data, the same evolutionary correction is applied to the absolute
magnitudes, $E(z) = q_0 (1 + q_1 (z - z_0)) (z-z_0)$, with $q_0=2$, $q_1=-1$ and $z_0=0.1$.
$\xi(r_p, \pi)$ is integrated using the same $\pi_\mathrm{max}=60~\hMpc$ as \citet{Zehavi2011}.
The lower panels show the relative clustering of the red and blue samples, compared to the total sample.
For the two fainter samples, we find reasonable agreement between the mock and the
SDSS clustering measurements. Using the new colour distributions has increased the clustering amplitude of the red galaxies, improving the
clustering of the mock compared to SDSS. For the brightest sample, there is a difference in the overall clustering amplitude, where the mock is more strongly clustered than the data. 
For this sample, the new colour distributions have little effect on the clustering
on large scales, while the colour-dependence is made stronger than is seen in the data on small scales.
We find that the relative clustering of the red galaxies in the lower panels is in good 
agreement with SDSS, and is improved by using the new colour distributions. However, the relative clustering of the bright blue galaxies on small scales is low compared to the data.

At higher redshifts, we compare the clustering in the mock with projected correlation function
measurements from GAMA \citep{Farrow2015}. Since the GAMA samples are defined based on cuts
in ${}^0M_r$ and ${}^{0}(g-r)$, the galaxies in the MXXL mock must be $k$-corrected to a new
reference redshift of $z_\mathrm{ref}=0$. This can be done using our original set of 
colour-dependent $k$-corrections. For the absolute $r$-band magnitudes, we convert from
$z_\mathrm{ref}=0.1$ to $z_\mathrm{ref}=0$ using
\begin{equation}
\label{eq:kcorr_mag_0}
{}^0M_r = \magr + \kcorrr(0),
\end{equation}
and similarly for the $g-r$ colours,
\begin{equation}
\label{eq:kcorr_col_0}
{}^{0}(g-r) = \gr + \kcorrg(0) - \kcorrr(0).
\end{equation}
To be consistent with the GAMA measurements, we also apply the same evolutionary correction,
$E(z) = Q(z-z_0)$, with $Q=1.45$ and $z_0=0$.
The projected correlation functions are shown in Fig.~\ref{fig:wp_colour_gama} in 3 redshift
bins, where in each bin the same magnitude cuts are used that are given in figure~14 of 
\citet{Farrow2015}. Galaxies are split into red and blue samples using a colour
cut ${}^{0}(g-r) = -0.030 ({}^0M_r - {}^0M_*) + 0.678$, where the values of 
${}^0M_*$ are given in table~1 of \citet{Farrow2015}. In the $0.14 < z < 0.24$ bin,
the amplitude of the clustering of the red galaxies has increased with the 
new colour distributions, bringing the mock into better agreement with the GAMA data,
particularly at small separations. The clustering of the blue galaxies is almost
unchanged. A similar effect is seen at intermediate redshifts, while in the highest
redshift bin, the new colour distributions have a very small effect on the 
clustering measurements. A difference in the shape of the projected correlation function
on large scales is also seen in this bin, compared to the GAMA data. 

The angular correlation function of galaxies in the MXXL mock is shown in
Fig.~\ref{fig:wtheta}, compared to measurements from BGS targets in the Legacy Imaging Survey, from BASS/MzLS \citep{Zarrouk2021}.\footnote{We only show the clustering from BASS/MzLS, but this is in good agreement with DECaLS.}
The upper panel shows the angular clustering in bins of apparent $r$-band
magnitude, with no cuts in redshift. The angular clustering of the faintest
samples is in very good agreement with the data, while for the brightest
samples, galaxies in the MXXL mock are slightly more strongly clustered.
This is similar to what was seen in the original MXXL mock \citep[see figure~11 of][]{Zarrouk2021}.
The lower panel shows the angular clustering for two of the samples, which have
been split by colour into red and blue galaxies. Here, the cut in colour is done
using the median $g-r$ colour for each sample. Again, we see good agreement
with the BASS/MzLS data. For the brightest sample, the colour dependence at
small scales is slightly stronger in the mock than is seen in the data.
We find that compared to the previous MXXL mock, there is very little change in the faint sample,
while the colour-dependence in the clustering of the brighter sample is slightly stronger with the
new colour implementation.

The MXXL mock, with colours assigned using fits to the GAMA colour-magnitude
diagram, is able to reproduce well a range of galaxy clustering statistics,
compared to measurements from SDSS, GAMA and the DESI legacy imaging
surveys, and is an improvement over the previous MXXL mock.

\section{Luminosity function: dependence on environment and colour}
\label{sec:lf_environment}

As an example application of using the MXXL mock catalogue, we investigate how the
luminosity function in the mock depends on galaxy colour and environment.
When constructing the mock, the HOD galaxies are assigned magnitudes to reproduce
an evolving luminosity function. This indirectly adds an environmental
dependence to the luminosity function, since the brightest galaxies reside in very dense environments in the most massive haloes, whereas 
the faint galaxies live in lower density environments. While the overall
luminosity function agrees with measurements from data, it is not guaranteed to
agree when split by environment. Similarly, the colour assignment is applied to
galaxies based on whether they are centrals or satellites. This adds colour dependence
to the luminosity function in the mock, which again is not guaranteed to match measurements from data.

\subsection{Local galaxy density}

The local density of a galaxy is calculated using a density defining population (DDP)
of galaxies \citep{Croton2005}. The DDP sample of galaxies is a volume limited sample, defined by an absolute magnitude
threshold and redshift range. A galaxy within the DDP sample would be observable at any redshift
within this range. 

We use the DDP1 sample of \citet{McNaught-Roberts2014}, which is for
galaxies with $0.039<z<0.263$ and $-21.8 < {}^{0}M_r^e < -20.1$, where the absolute magnitude has been 
$k$-corrected to a reference redshift $z_\mathrm{ref}=0$ (see Eq.~\ref{eq:kcorr_mag_0}), and an evolutionary correction applied. 
We use the same correction of $E(z) = Q_0(z-z_\mathrm{ref})$, where for all galaxies, $Q_0=0.97$.

The DDP sample can then be used to measure the local environment around each galaxy in the full sample, which covers the full magnitude range of the mock.
The number of DDP galaxies, $N_s$, is counted within a sphere of radius $r_s = 8~\hMpc$.
The local galaxy density within this sphere is
\begin{equation}
\rho = \frac{N_s}{\frac{4}{3}\pi r_s^3}.
\end{equation}
Since the full sample of galaxies in the mock is complete to $r=20.2$, and covers the full sky, 
we do not apply any completeness corrections. The redshift range of the full sample is made slightly smaller than the DDP sample to avoid incompleteness at these boundaries 
(we use $z_\mathrm{min}=0.059$ and $z_\mathrm{max}=0.243$).
The local overdensity is
\begin{equation}
\delta_8 = \frac{\rho-\bar{\rho}}{\bar{\rho}},
\end{equation}
where $\bar{\rho}$ is the mean density of DDP galaxies within the DDP volume 
(i.e. $N_\mathrm{DDP}/V$, where $N_\mathrm{DDP}$ is the total number of DDP galaxies, and $V$ is the volume of the DDP sample between $z_\mathrm{min}$ and $z_\mathrm{max}$)

\subsection{Luminosity function: environmental dependence}

The luminosity function is computed for galaxies in bins of $\delta_8$
using a $1/V_\mathrm{max}$ weighting, where $V_\mathrm{max}$ is the volume in which
it is possible to observe a galaxy with magnitude $\magr$ and colour $\gr$.
The luminosity function curves are then normalised to take into account the effective volume of the 
different density bins. The densest environments (with large $\delta_8$) only cover a very small
fraction of the volume, while the fraction covered by low density environments is
much larger. To calculate the fraction of the volume in each density bin, 
we generate a set of $N_r$ random galaxies which are uniformly distributed across the sky, but with
redshifts drawn from the galaxies in the mock. The overdensity, $\delta_8$, is calculated for the randoms using
the same set of DDP galaxies as before, and the luminosity function is weighted by $1/f_\delta$, where
\begin{equation}
f_\delta = \frac{N_{r,\delta}}{N_r}.
\label{eq:vol_fraction}
\end{equation}
Here, $N_r$ is the total number of randoms, and $N_{r,\delta}$, is the number of galaxies in the $\delta_8$ bin.

We compare the luminosity function in each density bin with a reference
luminosity function. We first measure the total luminosity function, $\phi_\mathrm{tot}$,
in the mock, for galaxies in the same redshift range. In each density bin,
the reference luminosity function is then
\begin{equation}
\phi_\mathrm{ref} = \frac{1+\langle\delta_8\rangle}{1+\langle\delta_{8,\mathrm{tot}}\rangle} \phi_\mathrm{tot},
\label{eq:lf_normalisation}
\end{equation}
where $\langle\delta_8\rangle$ is the mean value of $\delta_8$ in
each density bin, and $\langle\delta_{8,\mathrm{tot}}\rangle$, is a weighted mean
of the $\delta_8$ of all galaxies, taking into account the fraction of the volume occupied by galaxies with different values of $\delta_8$ \citep{McNaught-Roberts2014}.
This is given by
\begin{equation}
\langle\delta_{8,\mathrm{tot}}\rangle = \frac{ \sum_{i} \delta_{8,i} f_{\delta,i}}{\sum_{i} f_{\delta,i}} ,
\end{equation}
where the volume fraction, $f_{\delta,i}$, is evaluated from the randoms in narrow bins of density (using Eq.~\ref{eq:vol_fraction}). 
We obtain a value of $\langle\delta_{8,\mathrm{tot}}\rangle=0.026$, which is slightly
larger than the value of 0.007 found in \citet{McNaught-Roberts2014}. Since both numbers are close to zero, this has a negligible impact on the normalisation of the reference luminosity function.

\begin{figure} 
\centering
\includegraphics[width=\linewidth]{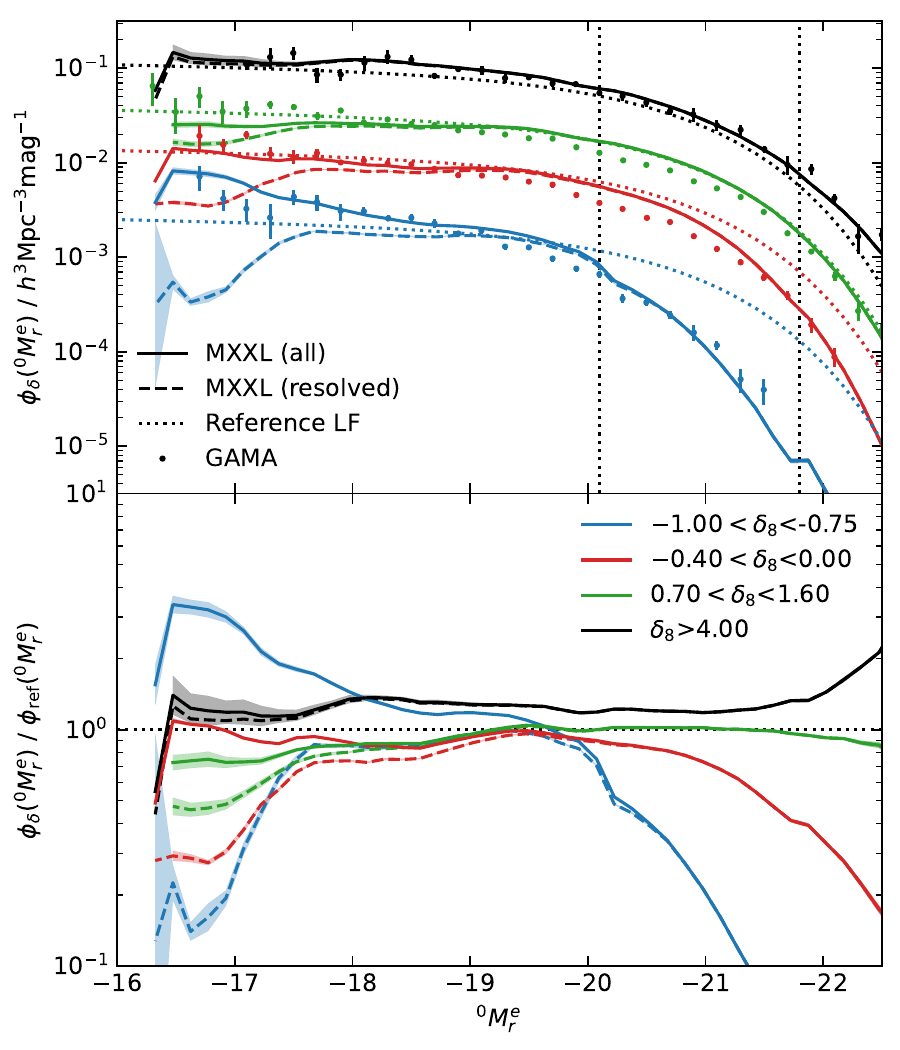}
\caption{\textit{Top panel}: environmental dependence of the luminosity function, in different bins
of $\delta_8$. The lowest density regions are shown in blue, with the highest density in black. The
solid curves show the luminosity function measured from all galaxies in the MXXL mock, while the
dashed curves only include galaxies that reside in resolved haloes. The shaded regions indicate
the error on the mean, from splitting the full sky into 9 regions. The dotted curves indicate the
reference luminosity function, which is a Schechter fit to the total luminosity function in the MXXL
mock, normalised according to Eq.~\ref{eq:lf_normalisation}. The luminosity function
measured in the GAMA survey is shown by the points with error bars. The magnitude range of
the DDP sample is shown by the vertical dotted lines.
\textit{Bottom panel}: Ratio with respect to the reference luminosity functions (dotted curves in the upper panel).}
\label{fig:lf_environment}
\end{figure}

The luminosity function in the mock split into 4 density bins is shown in the upper panel of Fig.~\ref{fig:lf_environment}, 
as in figure 6 of \citet{McNaught-Roberts2014}, where the highest density is shown in black, and lowest density
in blue. The solid curves are the measurements from the MXXL mock, and the points with error bars are from GAMA. These are in good agreement, particularly for the lowest and highest density samples, while for the
intermediate densities, there is a slight difference in the shape of the luminosity function at the knee.
The dotted line is the reference luminosity function.
The lower panel shows 
the ratio of each luminosity function to the reference. The shape of the different curves is very similar to what is
seen in the GAMA data in \citet{McNaught-Roberts2014}. At the bright end, there is a clear trend in shape with
density, where luminosity function falls off more rapidly with decreasing density. There are also differences
in shape at the faint end, with the lowest density luminosity function having a steeper slope.
The dashed lines show the effect of removing the unresolved haloes, which were positioned randomly in
the lightcone. Since they are randomly positioned, they are much more likely to be in low density regions,
greatly reducing the faint end slope for the blue curve (underdense regions), while having a negligible effect on the black curve (overdense regions).

\subsection{Luminosity function: colour dependence}

\begin{figure*} 
\centering
\includegraphics[width=\linewidth]{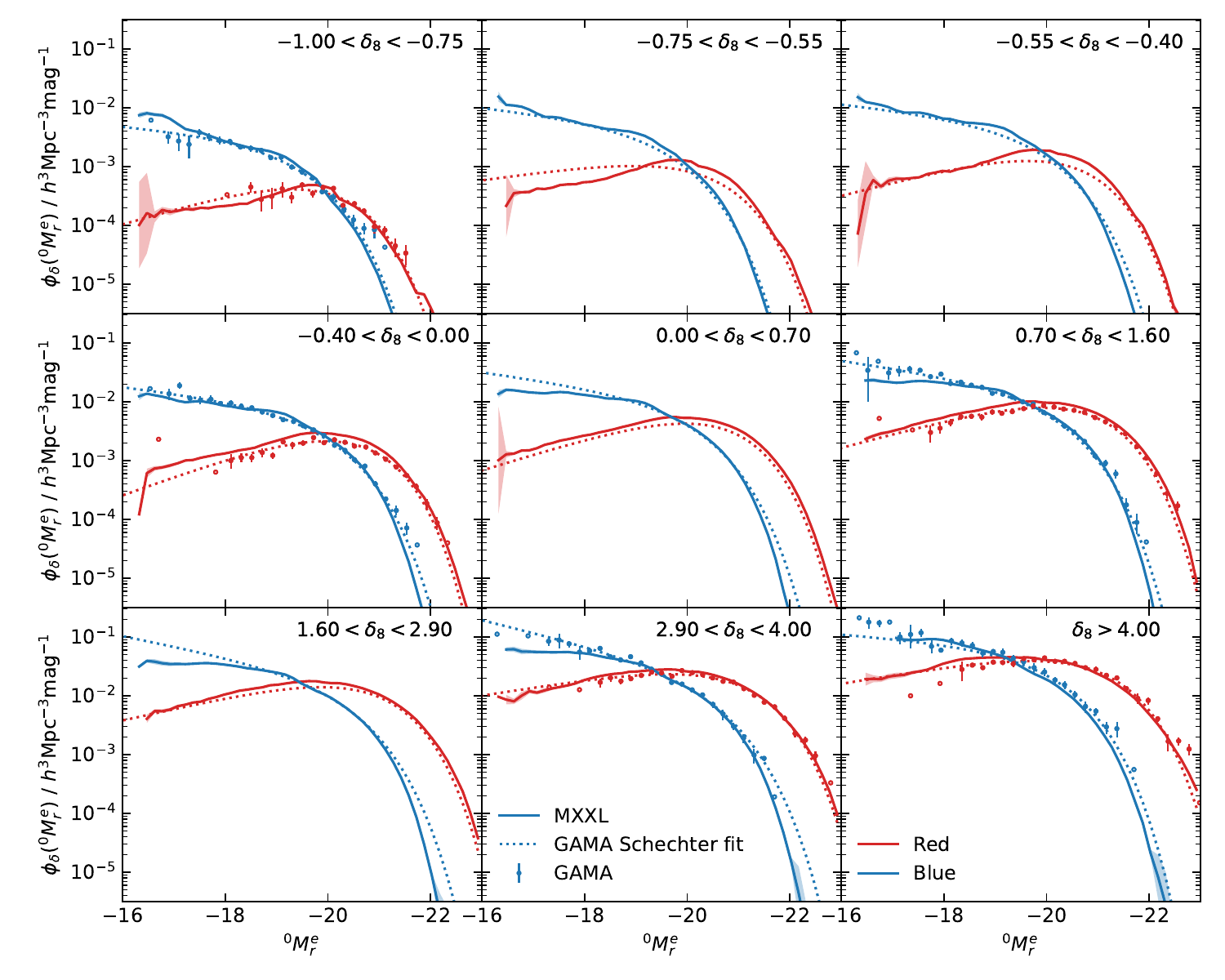}
\caption{Luminosity function in the MXXL mock, split by colour and environment. Each panel
is for a different bin in $\delta_8$, with the lowest density regions in the upper left panel, 
and highest densities in the lower right panel. Solid curves are the measurements from the 
MXXL mock, for the red and blue samples, plotted in red and blue, respectively. 
Points with error bars are the luminosity function measurements from GAMA, taken from figures~8 and 12 of \citet{McNaught-Roberts2014}. Open circles indicate that the errors could not be reliably estimated.
Dotted curves are Schechter fits to the measurements from GAMA, where the Schechter parameters are taken from figure~13 of \citet{McNaught-Roberts2014}.}
\label{fig:lf_colour}
\end{figure*}

We also calculate the luminosity in different environments, split into red and blue
galaxies. This colour split is done with the cut ${}^0(g-r)=0.63$,
where we convert the rest-frame colours in the mock to the reference redshift
$z_\mathrm{ref}=0$ using Eq.~\ref{eq:kcorr_col_0}.
As in \citet{McNaught-Roberts2014}, the evolutionary correction applied to the absolute magnitudes
is different for the red and blue galaxies. This $E$-correction uses $Q_{0,\mathrm{red}}=0.80$ 
and $Q_{0,\mathrm{blue}}=2.12$.

The luminosity function, split by environment and colour, is shown in Fig.~\ref{fig:lf_colour}. The solid
curves are the measurements from the mocks, and the dotted lines are Schechter fits to the GAMA
data, with the lowest densities in the top left panel, and the highest densities in the bottom right.
The Schechter parameters are taken from figure~13 of \citet{McNaught-Roberts2014}.\footnote{The $\alpha$ parameters in figure~13 of \citet{McNaught-Roberts2014} are provided relative to a reference Schechter function, with $\alpha_\mathrm{tot,red}=-0.38$ and $\alpha_\mathrm{tot,blue}=-1.37$.}
The galaxies in the mock catalogue show good agreement with the trends in the GAMA data. 
In each panel, the slope at the faint end, $\alpha$, is negative for blue galaxies and positive
for red galaxies, and the characteristic magnitude, $M_*$, is smaller for the blue galaxies.
Bright, central galaxies, are more likely to be red, while faint satellite galaxies are more likely to be blue.
The agreement between mock and data, 
is remarkably good, given that the mock was not tuned to 
reproduce these quantities.

\section{Conclusions}
\label{sec:conclusions}

In this paper, we present an updated version of the MXXL mock galaxy catalogue, which
was originally described in \citet{Smith2017}. Several improvements have been made to the
mock to improve agreement with measurements from data.

We test different methods for interpolating haloes between snapshots to build a lightcone.
This includes cubic interpolation, which was used in the original MXXL mock, linear interpolation, where positions and velocities are interpolated independently,
a constant acceleration interpolation, which is like linear interpolation but 
with positions that are consistent with the velocities,
and finally no interpolation, using a single simulation snapshot at the median 
redshift of the galaxy sample.

We find that cubic interpolation leads to extreme velocities half way between snapshots.
There are differences in the set of particles identified as belonging to the halo
at each snapshot, which leads to inconsistencies in the positions and 
velocities of the halo. Linear interpolation and constant acceleration interpolation
do not have these issues. We see differences at small scales ($\sim 0.6~\hMpc$) compared
to a mock with no interpolation, built from a single snapshot. This is due to
our implementation of halo mergers when constructing the lightcone. In redshift space,
this difference is small, particularly when satellite haloes are also included. 
In the MXXL mock catalogue, we adopt linear interpolation.

The assignment of $\gr$ colours to galaxies in the original MXXL mock was done using a
parametrisation of the SDSS colour-magnitude diagram, which was modified to add evolution
and to better agree with the distributions measured in data from the GAMA survey.
While the rest-frame colour distributions in the original MXXL mock are in reasonable agreement with GAMA,
small differences are much more apparent in the observer-frame colour distributions,
particularly at faint magnitudes, close to the magnitude limit of $r=19.8$.
In order to improve the colour distributions in the mock, we fit a new set of colour
distributions directly to the GAMA data. The bimodal colour distribution, in each bin of magnitude and redshift, is
described well as a double Gaussian.
In several bins of redshift, we fit a broken linear function to the mean and rms of
the red and blue sequences, as a function of absolute magnitude, in addition to the 
fraction of galaxies that are blue. These fits also take into account incompleteness at the faint limit of GAMA. These new colour distributions are used when 
assigning colours to galaxies in the updated mock, interpolating between redshift bins.
Both the rest-frame and observer-frame colour distributions show good agreement with the GAMA data, which is greatly improved compared to the original mock, and also with BGS targets in the DESI
Legacy Imaging Surveys. We extrapolate the
colour distributions to magnitudes fainter than $r=19.8$, and the agreement with DESI
remains good.

We compare the colour-dependent clustering in the mock with measurements from SDSS and GAMA
at a range of redshifts. The new colour distributions improve the relative clustering of
the red and blue galaxies. We also compare the angular clustering with BGS targets from the Legacy Imaging Surveys. The overall clustering amplitude is slightly higher for the brightest
samples, and the colour-dependent clustering shows good agreement with the data.

We also investigate how the luminosity function of galaxies in the mock depends on environment and colour. 
Since the HODs used to construct the mock depend on halo mass, and colour assignment depends on magnitudes, an environmental dependence is indirectly added to the mocks, but it is not guaranteed that this matches the measurements from data.
We measure the local overdensity of galaxies by using a density defining population (DDP) of galaxies, and counting the number within a sphere around each galaxy of radius $8~\hMpc$. We find that the trend of the luminosity function with environment agrees well with GAMA, although there are some differences in the overall shape of the luminosity function at intermediate densities. 
When the luminosity function is also split by colour, the trends in the mock show good agreement with the GAMA measurements.

The mock only contains $r$-band magnitudes and $g-r$ colours. However, mock galaxies could be matched to BGS galaxies,
magnitudes
and colours in other bands to be assigned. In \citet{DongPaez2022}, this is done on SDSS
mocks built from the Uchuu simulation. By matching galaxies in the mock to the data, 
based on absolute magnitude, colour and redshift, 
the mock galaxies can be assigned magnitudes in other bands, stellar masses and star formation rates. This works well at reproducing the correct
distributions for these quantities in the mock.

\section*{Acknowledgements}

AS would like to thank Michael Wilson for measuring the colour distributions of the BGS targets, and Svyatoslav Trusov for identifying the issues with the velocities in the original MXXL mock.
SC and PN acknowledge the support of the STFC Consolidated Grant  ST/T000244/1.
This work used the DiRAC@Durham facility managed by the Institute for Computational Cosmology on behalf of the STFC DiRAC HPC Facility (www.dirac.ac.uk). The equipment was funded by BEIS capital funding via STFC capital grants ST/K00042X/1, ST/P002293/1, ST/R002371/1 and ST/S002502/1, Durham University and STFC operations grant ST/R000832/1. DiRAC is part of the National e-Infrastructure.
For the purpose of open access, the author has applied a Creative Commons Attribution (CC BY) licence to any Author Accepted Manuscript version arising from this submission.

\section*{Data Availability}

The MXXL mock catalogue underlying this article is available in the Millennium database at \url{http://icc.dur.ac.uk/data/}.



\bibliographystyle{mnras}
\bibliography{ref} 








\bsp	
\label{lastpage}
\end{document}